\documentclass[10pt,aps,prd,nofootinbib,superscriptaddress,twocolumn,preprintnumbers,balancelastpage]{revtex4-2}
\usepackage[colorlinks=true
,urlcolor=blue
,anchorcolor=blue
,citecolor=blue
,filecolor=blue
,linkcolor=blue
,menucolor=blue
,linktocpage=true
,pdfproducer=medialab
,pdfa=true
]{hyperref}
\usepackage{dcolumn}
\usepackage[dvips]{graphicx}
\usepackage{epsfig}
\usepackage{amsmath,amsfonts,amssymb,amsthm}
\usepackage{mathrsfs,mathtools}
\usepackage{verbatim}
\usepackage{psfrag}
\usepackage{bm}
\usepackage{bbm}
\usepackage[utf8]{inputenc}
\usepackage[dvipsnames]{xcolor}
\usepackage{slashed}
\usepackage{upgreek}
\usepackage{enumitem}
\usepackage{float}
\usepackage[T1]{fontenc}
\usepackage{soul}
\usepackage{pgfplots}
\usepackage{extarrows}
\pgfplotsset{compat=1.18}
\usetikzlibrary{external,decorations.pathmorphing}
\usepackage{cases}
\usepackage{empheq}
\usepackage{physics}

\begin{document}

\preprint{MITP-25-078}

\title{Searching for Ultralight Dark Matter with MOLeQuTE: a Massive Optically Levitated Quantum Tabletop Experiment}

\author{Louis Hamaide}
\altaffiliation{Corresponding author: hamaidel@uni-mainz.de}
 \affiliation{INFN Naples, Strada Comunale Cinthia, 80126 Naples, Italy}
 \affiliation{Institut f\"ur Physik, Johannes Gutenberg-Universit\"at Mainz, Staudingerweg 9, 55128 Mainz, Germany}
\author{Hannah Banks}%
\affiliation{DAMTP, Cambridge University, Wilberforce Rd, Cambridge CB3 0WA, UK}%
\affiliation{Center for Cosmology and Particle Physics, Department of Physics, New York University, New York, NY 10003, USA}
\author{Peter Barker}
 \affiliation{Department of Physics and Astronomy, University College London,
Gower Street, London WC1E 6BT, United Kingdom}
\author{Andrew A. Geraci}
 \affiliation{Center for Fundamental Physics and Center for Interdisciplinary Exploration and Research in Astrophysics, Department of Physics and Astronomy, Northwestern University, Evanston, Illinois 60208, USA}


\date{\today}

\begin{abstract}

Many well theoretically motivated models of ultralight dark matter are expected to give rise to feeble oscillatory forces on macroscopic objects. Optically trapped sensors have high force sensitivities but have remained relatively unexplored in this context. In this work we propose a new, tunable, optically trapped sensor specifically designed to detect such forces. Our design features a high-mass ($\sim$ mg) plate whose high aspect ratio allows its weight to be supported by a vertical beam without excessive heating. We present the first systematic analysis and optimisation of quantum noises in optically trapped systems and show that our setup has the potential to operate at the standard quantum limit with current off-the-shelf technologies. We demonstrate that our sensor could offer unique access to large regions of uncharted parameter space of vector B-L dark matter, with projected sensitivities that could advance existing limits by several orders of magnitude over a broad range of frequencies.

\end{abstract}

\maketitle

\section{Introduction\label{sec:intro}}

Identifying the fundamental nature and properties of the Dark Matter (DM) understood to harbour $\sim 25\%$ of the energy budget of the Universe, remains one of the greatest open challenges for contemporary science. Over the past few decades, experimental attention has largely been directed towards so-called `Weakly Interacting Massive Particles' (WIMPs) and adjacent models with $\sim$ GeV  masses and weak scale couplings to the Standard Model (SM). Despite being the subject of intense scrutiny from a wide variety of direct, indirect and collider-based searches, no conclusive evidence for such particles has yet been attained \cite{PDG,Cirelli:2024ssz}. Whilst there remains interesting parameter space in this vicinity yet to explore, these null results have served to motivate a broader exploration of the vast landscape of theoretical possibilities open to DM \cite{Battaglieri:2017aum}. In particular, recent transformative advances in quantum sensing technologies have opened the window to so-called ultralight dark matter (ULDM) comprising sub $\sim$ eV mass particles with incredibly feeble couplings to the SM \cite{Jaeckel:2010ni,Antypas:2022asj}. Such fields are necessarily bosonic and, in order to saturate the local DM abundance, have extremely high phase space densities. As a consequence they present as a persistent, coherently oscillating classical wave over the scales relevant to laboratory and astrophysical searches, giving them starkly different phenomenology and hence  detection opportunities to heavier particle-like DM candidates. Whereas searches for WIMP-like DM typically seek impulses resulting from individual scattering events, direct detection of ULDM demands a distinct set of techniques which aim to exploit the spatial and temporal coherence of the DM field \cite{Battaglieri:2017aum,Antypas:2022asj,Jaeckel:2022kwg}.

A number of highly motivated DM candidates fall within this class including the QCD axion and other pseudoscalar axion-like particles (ALPs) \cite{PhysRevLett.38.1440,Peccei:1977ur,Weinberg:1977ma,Wilczek:1977pj,Shifman:1979if,Dine:1981rt,Zhitnitsky:1980tq,Kim:1979if,Preskill:1982cy,Abbott:1982af,Dine:1982ah,Svrcek:2006yi,Arvanitaki:2009fg,SHIFMAN1980493,DiLuzio:2020wdo,Choi:2020rgn}, spin-1 vectors such as the dark photon \cite{HOLDOM1986196,FAYET1990743,Fabbrichesi:2020wbt,Abel:2008ai,Goodsell:2009xc,Nelson:2011sf,Arias:2012az,Graham:2015rva,Agrawal:2018vin,Bastero-Gil:2018uel,Dror:2018pdh,Co:2018lka,Caputo:2021eaa,Adshead:2023qiw,Cyncynates:2023zwj,Cyncynates:2024yxm} and spin-0 scalars including moduli \cite{Dimopoulos:1996kp,Arkani-Hamed:1999lsd,Burgess:2010sy,Cicoli:2011yy}, dilatons \cite{Damour_1994,TAYLOR1988450}, Higgs Portal DM \cite{Piazza:2010ye} and the relaxion \cite{Banerjee:2018xmn}. The phenomenology, and in turn detection opportunities for a given model of ULDM ultimately depends on the way in which it couples to the SM. Different couplings produce distinct effects on the visible sector ranging from the apparent variation of fundamental constants, to oscillating torques and forces on macroscopic objects. In this work we specifically focus on those couplings which manifest as an oscillatory force on SM particles, with a frequency set by the mass of the DM field. These arise both in scalar DM models with an effective coupling to nucleons, as well as spin-1 vector bosons of a gauged U(1) symmetry which conserves baryon minus lepton (B-L) number \cite{Fayet:1980ad,Fayet:1980rr}. We adopt these models, which shall be formally introduced in Sec.~\ref{sec:dm_couplings}, as the theoretical benchmarks to illustrate this study.

A wide variety of increasingly competitive detection schemes have been proposed or built to search for ULDM-induced forces to date, many of which leverage
high-sensitivity, low-exposure quantum sensors which operate at, or close to, the quantum limit (see e.g.~Refs.~\cite{Badurina:2019hst,Arvanitaki:2014faa,Amaral:2024rbj,Kalia:2024eml,Carney:2019cio,Higgins:2023gwq,Deshpande:2024bul,Li:2023wcb,Manley:2020mjq,Manley:2019vxy,chowdhury2025optomechanicalaccelerometersearchultralight} for a non-extensive overview). Of these, mechanical sensors -  in essence harmonic oscillators - have shown particular promise, offering sensitivity to oscillating external forces acting on the system close to its mechanical (resonant) frequency. A wide variety of mechanical resonators have been developed, including clamped solid state oscillators, membranes and pendulums \cite{Tsaturyan:2016icv,PhysRevApplied.14.014041,Aspelmeyer:2013lha,PhysRevX.6.021001,Matsumoto:2018via}. These devices are coupled to optical fields to read out their displacement and velocity, from which any external force (e.g.~from ULDM) can be inferred. Of these systems, suspended cavity mirrors acting as pendulums have been a popular choice for fundamental physics investigations, with applications ranging from the search for new particles to gravitational wave astronomy \cite{Carney:2019cio,Geraci_2019,LIGOScientific:2007fwp}. In a proof of concept study, Ref.~\cite{Carney:2019cio} demonstrated that suspended cavity mirrors have the potential to significantly extend the reach of existing experiments to ULDM-induced forces. Although these systems lose tunability of their natural mechanical frequency (achieved by adjusting the suspension string length) at high frequencies $>\mathcal{O}(100$~Hz), they may be made tunable by exploiting the optical spring effect \cite{Aspelmeyer:2013lha}. A consequence of the optical spring effect however is that the mechanical frequency becomes dependent on the laser power such that the freedom to tune the power at an arbitrary resonant frequency (e.g.~to optimise the quantum noises) is lost. Tuning the power at a fixed mechanical frequency is  commonly touted as a means to minimise the quantum noises to reach the `Standard Quantum Limit' (SQL) which is ideally achieved`on resonance' i.e.~at the mechanical frequency \cite{Aspelmeyer:2013lha,Bowen_Milburn}. This motivates the search for complementary solutions where 1) the mechanical resonance is tunable over a large range of frequencies and 2) the power can be independently varied to minimise the intrinsic quantum noises on resonance. While the setup proposed in this work does not fully satisfy these criteria, it does possess a mechanical resonance that can be tuned over several orders of magnitude via both the laser power and the distance to the laser focus. It is also  able to reach the SQL at frequencies substantially closer to the mechanical resonance than other optical systems such as suspended cavity mirrors \cite{Carney:2019cio,Matsumoto:2018via}. Although this proximity does not itself improve the sensitivity (the benefit is only achieved exactly on resonance), the design principles underlying our setup may prove key to the development of tunable mechanical resonators capable of achieving the SQL on resonance in the future.

An alternative to mechanical suspension or optical spring effects which has been more recently explored is to `trap' the sensor in a harmonic potential created with electromagnetic (henceforth referred to  as `optical'), electric or magnetic fields, or combinations thereof \cite{Higgins:2023gwq,Latorre_2022,Biercuk:2010tte,Ivanov_2016,ji2025levitatedsensormagnetometryambient,Millen_2020,Jain:2016,Magrini:2020agy,Moore_2021,ASHKIN1992569,Ashkin_review:1997,Gosling:2023lgh,Monteiro:2020qiz}. In this work, we focus specifically on optical trapping for which the calculation of quantum noises is generally more established (see e.g.~Refs.~\cite{Millen_2020,Magrini:2020agy,Jain:2016,Gosling:2023lgh} for a non-extensive overview) than for the latter options, making such devices  more amenable to a systematic study. 
The use of optical fields for simultaneous trapping and readout was pioneered in levitated optomechanics by Refs.~\cite{Li2011,Jain:2016}. Such systems have been shown to have a wide variety of applications \cite{Winstone:2023whl} ranging from demonstrating the ground state cooling of macroscopic objects \cite{Tebbenjohanns2021,delic2020}, to testing wavefunction collapse models \cite{Vinante_2019}.  They can also be used, analagously to suspended setups, for force sensing \cite{Monteiro:2020qiz,Gosling:2023lgh}. Optically trapped systems offer excellent control over both translational and rotational degrees of freedom, and retain tunability up to remarkably high frequencies:  $\mathcal{O}(1~\text{MHz})$ for forces and $\mathcal{O}(1~\text{GHz})$ for torques respectively \cite{Winstone:2023whl,Ju_2023}. Combined with their low setup and running costs, these promising features make these systems an attractive option for force sensing. Despite this, the space of possible optically trapped experiments remains largely unexplored and at present, no attempt has been made to optimise or purposefully design setups for the specific signals relevant to ULDM searches. 

Theoretical and experimental studies of optically trapped systems have overwhelmingly focused on trapping of objects of size tens of $~\mu\text{m}$ or smaller to date \cite{Moore_2021,Winstone:2023whl,Millen_2020}. Procedures for optically levitating nanoparticles are now well established and can achieve state-of-the-art noise power spectra of $S_\text{FF}\lesssim\mathcal{O}(10^{-40}~\text{N}^2/\text{Hz})$ \cite{Winstone:2023whl}. Given that the ULDM-induced force acting on the sensor scales with the sensor size (or equivalently mass), levitated nanoparticles experience comparatively low forces from background ULDM fields relative to competing classes of mechanical sensor such as suspended systems where vastly heavier objects with $\mathcal{O}(1~\rm mg)$ masses  are routinely used. For nanoparticles, provided that the characteristic size of the sensor remains smaller than the wavelength of the trapping optical field (the so-called Rayleigh regime), the quantum noise intrinsic to the system scales with the sensor mass in the same way as the DM-induced force. As a result, varying the mass of the sensor is of no benefit to the experimental reach, fundamentally limiting the long-term potential and applicability of these systems for ULDM detection. If however, a substantially larger object whose characteristic size exceeds that of the trapping wavelength (thus entering the geometric optics regime) could be optically trapped, the scaling of quantum noise with mass becomes less severe such that pushing to higher sensor masses offers substantial gains in sensitivity. 

The levitation of high mass objects in the geometric optics regime brings very different considerations to those of Rayleigh scattering nanoparticles. While there has been some consideration of levitating large masses in the literature (see e.g.~Refs.~\cite{Michimura:2016urt,Guccione_2013}), little to no attention has been given to developing experimental configurations optimised for force sensing. In this work we directly address this void by detailing an optically trapped setup in which the sensor weight is supported by a vertical beam, allowing for the levitation of $\sim$~mg  masses and above. For conventional low aspect ratio sensors, absorption of light from the vertical beam is expected to lead to excessive heating above the mg scale, ultimately melting most candidate materials including silica. This arises because the laser power required for levitation scales with the sensor volume whereas the cooling power from blackbody radiation scales with surface area. To overcome this limitation and facilate the levation of heavier objects we thus propose a sensor with a flat plate-like geometry, whose mass can be artibrarily increased by scaling its lateral dimensions whilst leaving the thickness unchanged. At such high-aspect ratios, the  area approximately scales with the mass, allowing substantially heavier sensors to be optically levitated without an increase in temperature. In this work we therefore study the levitation of both a short term $\sim$~mg sensor and a medium term $\sim$~g scale sensor. By undertaking a detailed analysis of the  noises intrinsic to such an experiment, we optimise the remaining experimental parameters to maximise the signal-to-noise ratio (SNR) for the high-quality oscillating external forces relevant to ULDM searches. In particular we present the first consistent optimisation of the quantum noises relevant to optically trapped systems to date. This procedure differs significantly from e.g.~cavity noise optimisation since the mechanical frequency of the system  and the optical power are related and thus cannot be independently varied. We show that optically trapped systems perform better when operating at  SQL off resonance as opposed to being on resonance but above SQL. The SQL noise we derive matches the off-resonance SQL noise of non-optically trapped sensors despite the different underlying forms of  noise. Using this optimised noise, we demonstrate that our setup has the potential improve upon the sensitivity of existing experiments to vector B-L ULDM, relying only on existing technologies and moderate resources.

This paper is organised as follows. In Sec.~\ref{sec:dm_couplings} we introduce the DM models that will form the theoretical targets for this study, and the spectrum of forces that they would induce on a macroscopic test mass. The proposed setup is then discussed in  Sec.~\ref{sec:setup}. In Sec.~\ref{ssec:quantum_noises} we outline the quantum noises intrinsic to optomechanical systems and detail, for the first time, how to optimise these specifically for optically trapped systems. We then comment on the relevant classical noises and discuss methods for mitigating their impact on the experimental sensitivity in Sec.~\ref{ssec:th_noises}. Finally,  in Sec.~\ref{sec:results} we provide sensitivity forecasts to vector B-L and scalar ULDM and discuss near-future and long-term prospects to fully exploit our setup with reasonable resources.

 We adopt natural units in which $\hbar = c = 1$ throughout this work except for when giving numerical results or when explicitly stated otherwise. We generically use $f$ and $\omega = 2 \pi f$ to denote  frequencies and the corresponding angular frequencies respectively.

\section{Forces from ULDM\label{sec:dm_couplings}}

In this section we briefly review the primary features of ULDM and detail some examples of concrete, well-motivated models whose couplings to the SM result in additional forces on macroscopic test masses. 

The term ultralight dark matter is generically used to refer to a bosonic field of mass $m_{\rm DM }$ $\lesssim 1$eV which is assumed to comprise an $\mathcal{O}(1)$ fraction of the local DM abundance. These particles have a de Broglie wavelength of $\lambda_{\rm DM} = 1 / (m_{\rm DM} v_{\rm DM})$, where $v_{\rm DM}$ denotes the root mean squared velocity of the DM particles. Assuming DM is virialised in the local galaxy, the virial theorem dictates $v_{\rm DM}$ to be $\sim 10^5$~m/s  ($\sim 10^{-3}$ in natural units).  Given the local DM abundance
$\rho_{\rm DM} \sim 0.4 $~GeV/cm$^3$ (assuming the Standard Halo Model \cite{Read_2014,Riccardo_Catena_2010}), the number of particles in a de Broglie volume $\lambda_{\rm DM}^3$  (i.e.~the phase space occupancy), is
\begin{equation}
N_{\rm db} = \frac{\rho_{\rm DM}}{m^4_{\rm DM}v_{\rm DM}^3} \sim 10^{3} \left(\frac{1 \rm eV}{m_{\rm DM}}\right)^4~,
\end{equation}
which, for sub-eV mass particles, is thus extremely high. As a result, quantum fluctuations are suppressed and the DM can be treated as a persistent classical field formed from the superposition of different plane waves with random phases and velocities $\bf{v}_\text{DM}$ that follow a Maxwell-Boltzmann distribution. To be cold, ULDM is necessarily produced non-thermally; we will comment further on specific mechanisms relevant to the models being studied below. 

Within each de Broglie volume, the DM field oscillates coherently with a Compton (angular) frequency of $\omega_{\rm DM} \sim m_{\rm DM}$. As a result of the finite velocity distribution, there is a small frequency spread in each volume of $\Delta \omega_{\rm DM} \sim v_{\rm DM}/\lambda_\text{DM}$. The different de Broglie volumes, which delineate distinct coherent spatial regions
composed of fields with velocity standard deviation $v_\text{DM}$,
remain temporally coherent over a coherence time
\begin{equation}
T_{\rm coh} = \frac{2 \pi}{\Delta \omega_{\rm DM}} \sim 10^6\left(\frac{2\pi}{\omega_{\rm DM }}\right)~.
\end{equation} 
The factor of $v_{\rm DM}^{-2}\approx 10^6$ relating $T_{\rm coh }$ and $(2\pi/\omega_{\rm DM})$ defines the quality factor $Q$ of the ULDM signal. Note that this is independent of the DM mass. 
The phase interference of different $k$-modes composing the ULDM field can further cause the total DM field amplitude to deviate from that  implied by the local DM density. The observed DM wave amplitude thus varies across coherent times, taking values which follow a Rayleigh distribution with mean given by the local DM density \cite{Foster_2021}. The effect of this on the statistical significance of exclusion bounds is typically captured by an $\mathcal{O}(1)$ factor, as discussed in e.g.~Refs.~\cite{evans2018shmrefinementstandardhalo,GREEN_2003,amaral2024vectorwavedarkmatter}. Given this, we will not consider this correction in this work. We will also ignore the bias in DM velocities coming from the Sun's motion around the Milky Way (the DM `wind') and the daily/annual variation in signal due to the Earth's rotation and its motion around the Sun. As we will see, our sensor exhibits the potential for directional force sensing which, given these effects, may offer the means to further enhance our sensitivity to ULDM if exploited in the future. Finally we neglect the effect of subdominant local dark matter overdensities/underdensities. A more careful treatment of all of these effects  can be found in Refs.~\cite{evans2018shmrefinementstandardhalo,GREEN_2003,amaral2024vectorwavedarkmatter,Baxter_2021}. We emphasise that the inclusion of  these effects is not expected to result in a significant change to our sensitivity forecasts and we therefore leave a more detailed assessment of their potential impact to future work.

ULDM can give rise to a variety of effects on visible matter, depending on the way in which it couples to the SM at the microscopic level. Over timescales shorter than $T_{\rm coh}$ these effects are characteristically oscillatory in nature with (in the case of a linear coupling) an angular frequency of $\omega_{\rm DM}$ and a Doppler broadened linewidth of $\Delta \omega_{\rm DM}$. In this work we are specifically interested in those couplings which manifest as an oscillatory force on macroscopic objects, i.e.~a levitated sensor, driving displacements of its centre of mass position relative to some reference (in our case the optical components forming the trap). Whilst there are numerous microscopic couplings which can give rise to such behaviour, to illustrate the capabilities of the experiment we focus on a couple of specific examples. Although our treatment here is not exhaustive, it serves to highlight the potential of our design as a novel probe of oscillatory DM-induced forces, paving the way towards more comprehensive phenomenological studies in the future. 

As we shall see, whilst these forces scale with the mass of the sensor, provided that the DM couples differently to protons, neutrons or electrons (as is the case for the models of interest) the resulting force is material dependent. This property is crucial if the force is to be detected directly; with composition-independent forces there would be no relative motion between the sensor and its position reference such that only force \textit{gradients} could be measured. As a result of this property, objects with different proton-to-neutron ratios (in our case the sensor and a physical reference frame such as an optical table) would be subject to a differential acceleration in the presence of the background DM field and thus move relative to each other. Such composition-dependent forces violate the weak equivalence principle and are subject to constraints from both torsion balance experiments~\cite{Wagner_2012} and MICROSCOPE~\cite{Berg_2018,MICROSCOPE:2022doy} which probe this.

\subsection{B-L coupled ULDM}

The first example is vector DM which couples directly to the SM via a gauge coupling. In this instance the ULDM field corresponds to the spin-1 gauge boson of an additional gauged U(1) symmetry, which interacts directly with the SM fermions via a coupling to the conserved gauge current. Within the SM there are a limited number of conserved currents that DM can couple to without introducing additional gauge anomalies\footnote{Models coupling to baryon number B are often considered, but are not anomaly-free \cite{Lee_2002}.} - one being baryon minus lepton number (B-L), which we shall henceforth focus on. Various different non-thermal  production mechanisms for vector ULDM have been proposed in literature. For the mass range of most relevance to our proposal, viable mechanisms include misalignment \cite{Arias:2012az,Nelson:2011sf}, the decay of cosmic string networks \cite{Long:2019lwl,Kitajima:2022lre} or by first populating an auxiliary sector (typically a scalar condensate) which then transfers energy to the vector boson via either parametric resonance or tachyonic instability~\cite{Co:2018lka,Agrawal:2018vin,Bastero-Gil:2018uel,Dror:2018pdh,Co:2021rhi}. For masses exceeding $\sim 10^{-5}$~eV, gravitational production from inflationary fluctuations~\cite{Graham:2015rva,Kolb:2023ydq} is also possible.

Writing the spin-1 gauge boson as $A'_{\mu}$, the Lagrangian density of such a theory contains an interaction term
\begin{equation}
\mathcal{L} \supset i g_\text{B-L} A'_{\mu} \overline{n} \gamma^\mu n~,
\end{equation}
where $g_\text{B-L}$ is a coupling strength and $n$ here is used to denote the neutron field. Following an analogous derivation to that of the Lorentz force in electromagnetism, this interaction gives rise to a force on macroscopic objects. The total DM force depends on the expression for the `electric field' component of $A'_{\mu}(\bf{x},t)$. The full expression sums over sine wave contributions each containing different DM velocities which appear as $\bf{k}\cdot\bf{x}$ where ${\bf k}= m_{\rm DM}{\bf v}_{\rm DM}$ with ${\bf v}_{\rm DM}$ having isotropically distributed angles and Maxwell-Boltzmann distributed amplitudes.  At a given point in space however we may approximate the background DM field as a single (monochromatic) sine wave over timescales shorter than the coherence time $T_\text{coh}$. In Fourier space, this corresponds to the case where the signal linewidth $\Delta \omega_{\rm DM}$ is equal to or smaller than the frequency resolution of the experiment (which is set by the inverse of the total time over which measurements are made).  The spatial dependence of the force can also be dropped since, at the Compton frequencies of interest, the wavelength of the DM field greatly exceeds typical laboratory scales (e.g.~the dimensions of the sensor). Assuming the polarization distribution of $A'_\mu$ is isotropic~\cite{amaral2024vectorwavedarkmatter}, the magnitude of the force in any one given direction can be written as
\begin{equation}
\label{eq:force_B-L}
\text{F}(t) =g_{\rm B-L} N_{\rm B-L}  \sqrt{\frac{2 \rho_{\rm DM}}{3}} \sin{(\omega_{\rm DM} t + \varphi_0)}~,
\end{equation}
where the phase $\varphi_0$ is arbitrary for each coherence time. Here, $N_\text{B-L}$ refers to the B-L charge of the test object, which for charge-neutral materials corresponds to the number of neutrons
$N_n = m(1 -(Z_s/A_s))/m_n$, where $m$ and $Z_s/A_s$ correspond to the test object mass and proton-to-nucleon ratio respectively, and $m_n$ is the mass of the neutron.

\subsection{Scalar ULDM}

Spin-0 (scalar) ULDM can also give rise to oscillatory forces on macroscopic objects. Many well-motivated DM candidates fall into this class including dilatons, moduli, higgs portal DM and the relaxion. Such scalar fields generically couple to the masses of SM fermions and gauge kinetic operators (e.g.~of photons and gluons) via  dimension-5 operators \cite{Damour_1994,Damour_2010} and can be produced  non-thermally via standard misalignment (as for axions) \cite{Preskill:1982cy,Abbott:1982af,Dine:1982ah,Antypas:2022asj} or thermal misalignment \cite{Batell:2021ofv,Piazza:2010ye,Banerjee:2018xmn}.  Given that the experiment takes place well below the QCD scale, protons and neutrons supplant quarks and gluons as the appropriate degrees of freedom to consider here. Bounds on couplings to protons or neutrons derived from low energy experiments can be straightforwardly translated to constraints on the fundamental couplings to quarks and gluons using the relations in e.g.~Refs.~\cite{Damour_1994,Damour_2010}. 

For the purpose of example, we focus on the force generated by a coupling to neutrons but note that this discussion can be straightforwardly generalised to couplings to  protons, electrons or combinations thereof, and the resulting constraints recast to any given model of interest. The leading-order non-derivative interaction between a generic scalar field $\varphi$ and neutrons that can be written down is
\begin{equation}
\label{eq:scalarlag}
\mathcal{L} \supset y_n \varphi \overline{n} n ~,
\end{equation}
where $y_n$ is a dimensionless coupling constant. This interaction  causes the neutron mass to appear to oscillate in time. It also generates a force on test objects from the gradient of the interaction potential. Assuming the DM velocity distribution is isotropic, the magnitude of the force in any one given direction is
\begin{equation}\label{eq:force_scalar}
\text{F}(t) = y_n v_{\rm DM}\, N_n \sqrt{\frac{2 \rho_{\rm DM}}{3}} \sin(\omega_{\rm DM} t + \varphi_0) ~,
\end{equation}
where we have once again assumed the DM to be a monochromatic plane wave. As can be seen, this is structurally analogous to the B-L case, albeit with a velocity-suppressed effective coupling. We note that as before, this force is crucially material dependent, and the associated acceleration is again dependent on the material's neutron-to-nucleon (or equivalently proton-to-nucleon) ratio. Whilst we will primarily focus on determining the sensitivity of our setup to vector B-L DM in this work, the analysis for the scalar couplings discussed in this section is identical (up to a constant factor of the DM virial velocity).

\section{Novel Geometric Optic Trap Setup}\label{sec:setup}

\subsection{Experimental Setup and Force Sensing}
\begin{figure*}[t!]
    \centering
    \includegraphics[width=1.98\columnwidth]{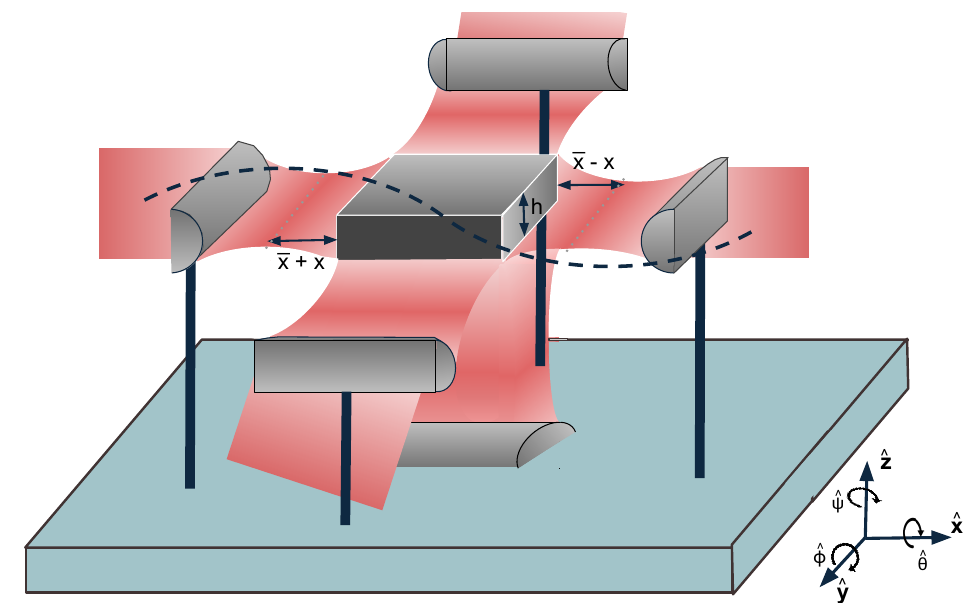}
    \caption{Conceptual perspective view of our proposed setup for sensing forces from dark matter (dashed line) using a silica plate (sensor) with a high-reflectivity coating. The sensor is levitated with a (vertical) optical beam and its motion along $\hat{\bf{x}}$ and $\hat{\bf{y}}$ is trapped independently via pairs of counterpropagating elliptical lasers which are focused using cylindrical lenses. Shown here is the final stage (blue-grey suspended plate) of the seismic isolation system to which the lenses are rigidly fixed. The perceived motion of the sensor is that relative to the plate. Not shown here are the photodetectors and/or windows placed around the setup to extract non-reflected light from our system. The distances $\bar{x}\pm x$ of each surface to the nearest focus (dotted lines in the $\hat{\bf{y}}$-direction) vary around the average $\bar{x}$. The sensor height  $h$ is also shown. Rotation directions of the sensor around the $(\hat{\bf{x}},\hat{\bf{y}},\hat{\bf{z}})$-axes away from the orientation shown here are denoted $(\hat{\bf\theta},\hat{\bf \phi},\hat{\bf{\psi}})$.}
    \label{fig:setup_diagram}
\end{figure*}

The proposed experiment comprises a large central mass - the sensor - which is optically trapped in both the $\hat{\bf x}$ and $\hat{\bf y}$ directions using pairs of counter-propagating elliptical Gaussian laser beams of wavelength $\lambda_L=1064$~nm directed along these axes as shown in  Fig.~\ref{fig:setup_diagram}.  Each beam is focused in the vertical direction with a  high numerical aperture (NA) cylindrical lens, which we take to be $\text{NA}\approx0.8$. The trapping beams reflect off the sensor which consists of a thin rectangular plate made of fused silica covered in a high-reflectivity coating. The coating is assumed to be a Bragg mirror formed from successive layers of silica and tantalum with a reflection coefficient $R\gtrsim 99.98\%$~\cite{Guccione_2013}, and has a total width of 2~$\mu$m. The net dimensions of the sensor (including both the substrate and coating) are  $3.5~\text{mm}\times 3.5~\text{mm} \times 5~\mu\text{m}$, corresponding to a total mass of $m\approx 0.2$~mg. Such a thin device can be fabricated by starting from a 3.5 mm square silica substrate of larger thickness (e.g.~1 mm) that is polished on each surface, coating the sides and bottom of the structure with a dielectric multilayer high reflectivity coating using electron beam evaporation or pulsed laser deposition, and then polishing the thickness down (from above) to the desired 5~$\mu$m thickness. Note that the top face of the sensor does not require coating.

To read off the position of the sensor (relative to the optical components), the light reflected off the coating is recaptured by the cylindrical lenses used to focus the incoming beams. The light of the reflected beam is  separated from the incoming optical field using a polarising beam splitter followed by a quarter-wave plate. The polarisation of the reflected light that re-enters the polarising beam splitter is thus rotated by 90 degrees with respect to that of the incoming light such that the two separate. The reflected light can then be used in a balanced homodyne measurement scheme, which is routinely deployed in optomechanical sensing devices e.g.~see Refs.~\cite{Magrini:2020agy,Gosling:2023lgh}. Any light from the horizontal beams that does not scatter off the sensor, is captured and removed by photodectors positioned above and below the plane of the sensor.

The optical components of the trap are rigidly attached to a plate which serves as an optical table. This plate forms the final stage of a system of nested oscillators (suspended plates) intended to damp motion induced by seismic vibrations of the external environment. We base our design on the  system described and used in Ref.~\cite{Geraci_2019}, consisting of three copper stages with the last one (the `cold stage') lowered into a cryogenic environment which is cooled to $\sim$ 4 K using liquid helium. To improve the seismic isolation performance further and obtain performances similar to current state-of-the-art seismic isolation systems, we plan to add an additional (fourth) stage. 

The choice of material used for the final stage plate is crucial to the performance of the experiment, as we now discuss. The final stage of the seismic isolation system is where the optical components are placed, and constitutes the reference frame with respect to which we measure the sensor's position and thus infer the force on the sensor. As we saw in Sec.~\ref{sec:dm_couplings}, both of the DM models of interest effectively couple to neutrons. Provided the sensor and support plate are composed of materials with different neutron-to-nucleon ratios they will thus experience a differential acceleration
\begin{align}
\label{eq:acc}
 a_{\rm rel} &= \left(\frac{Z_{\rm Si}}{A_{\rm Si}}-\frac{Z_{\rm p}}{A_{\rm p}}\right) a_0 \sin(\omega_{\rm DM} t + \varphi_0) 
\end{align}
where $a_0 = (\tilde{g}/m_n)\sqrt{2 \rho_{\rm DM}/3}  $ with $\tilde{g}$ denoting $g_\text{B-L}$ or $y_n v_{\rm DM}$ as appropriate for vector B-L ULDM and scalar ULDM  coupling solely to neutrons, respectively. In Eq.~\ref{eq:acc} we have used that the sensor is primarily composed of silica (Si) and use the subscript $\rm p$ to refer to the material of the support plate. This differential acceleration drives a relative displacement between the sensor and support plate. The force on the sensor that is inferred from measurements of the sensor position (relative to the plate) is thus $\widetilde{F} = m a_{\rm rel}$, where $m$ is the sensor's mass. 

Given the neutron-to-proton ratio of silica is 1:1, to maximise the relative displacement of the sensor and plate and in turn the sensitivity to DM, the neutron-to-proton ratio of the material used for final stage plate material should be as high as possible (within experimental constraints). For this reason, we select tungsten-186 for the final stage plate. This is both easily sourced and suitable for use in ultrahigh vacuum. With this choice, the difference in the proton-to-nucleon numbers of the sensor and the plate is 0.1.

\subsection{Optical Trapping Forces}

In this work we focus exclusively on sensing forces along the $\hat{\bf{x}}$-axis and are thus principally concerned with the trapping and control of the sensor in this direction. Trapping is provided by the beams directed along $\pm\hat{\bf{x}}$ which create a harmonic oscillator potential for the sensor. To see this we start by defining the average distance between the focus of the beam (which is a line in the $\hat{\bf{y}}$-direction) and the nearest surface of the sensor to be $\bar{x}$. The distance $\bar{x}$ corresponds to the equilibrium position of the sensor and is a design choice that can be tuned by moving one (or both) of the lenses  along the $\hat{\bf{x}}$-axis. We denote the Rayleigh range of the beams, i.e.~the distance from the beam waist to the point at which the spot size has increased by a factor of $\sqrt{2}$, as $x_r=\lambda_L/(\pi\text{NA}^2)$, where we remind the reader that NA is the numerical aperture. Provided $\bar{x}\gg x_r$ and $\bar{x}>h/\text{NA}$, where $h$ is the height of the sensor, the average intensity of the lensed beam at either surface is $I(\bar{x}) \approx \frac{I_0 x_r}{\bar{x}}$, where $I_0\propto P_L$
is the average beam intensity at the focus. Note that the intensity goes as $1/\bar{x}$ as the beam is only lensed in one direction. 

Given that the dimensions of the sensor exceed the laser wavelength, the interactions of the laser and sensor can be described using the geometric optics approximation. In this regime, the sensor reflects light off its surface as opposed to  behaving like a dielectric. As a result our sensor is necessarily trapped \textit{away} from the focus of the laser beams, in contrast to more commonly studied dielectric nanoparticles which are trapped at the beam's focus \cite{Millen_2020}.

Supposing the sensor is displaced from its equilibrium position by a small distance  $x$ in the $\hat{\bf x}$ direction, the net force on the sensor from light scattering off its sufaces  is \cite{Griffiths_2017} (see Appendix~\ref{app:trapping_potential} for details) 
\begin{equation}
F_{\textnormal{scatt}}(x) = -\frac{4I_0 x_rA}{\bar{x}^2}x \,\,\mathbf{\hat{x}}~,
\end{equation}
where $A$ denotes the cross-sectional area of the sensor normal to $\hat{\bf{x}}$. This describes a harmonic oscillator about the equilibrium position $\bar{x}$  with oscillation frequency 
\begin{equation}\label{eq:SHO_frequency}
\Omega = \sqrt{\frac{4 I_0 x_r A }{\bar{x}^2 m}} ~.
\end{equation}
For a fixed sensor geometry our system is highly tunable and benefits from a degeneracy between the laser intensity $I_0$ and $\bar{x}$ in that either one can be fixed whilst the other is tuned to obtain different values of $\Omega$ e.g.~in order to optimise the quantum noise for a specific dark matter search frequency. Note this is analogous to the degeneracy of tuning the laser power and NA when trapping nanoparticles \cite{Millen_2020}.

The weight of the sensor is countered using a vertical beam, also of $1064$~nm wavelength, directed upwards onto the underside of the sensor. The vertical beam is defocused in the $\hat{\bf{x}}$-direction and focused along the $\hat{\bf{y}}$-direction using cylindrical lenses similarly to the counter-propagating beams used to trap the sensor in the $\hat{\bf{x}}$-direction. The light from this beam reflects off the coating back into the cylindrical lens below, from which it is directed out of the system. This prevents it from being absorbed by the  vacuum chamber walls and heating the system. Most of the light can be recirculated to lower the input laser power requirements, while a small portion will need to be read out with a photodetector to ensure damping feedback loops can be implemented for control and stabilization of the sensor in the $\hat{\bf{z}}$-direction (for damping requirements see Sec.~\ref{sec:noises}).  An alternative to levitating the sensor would be to suspend it using a silica fibre string.  Although suspension systems have been more extensively studied \cite{Carney:2019cio,Deshpande:2024bul,Matsumoto:2018via,Aspelmeyer:2013lha,amaral2024vectorwavedarkmatter,LIGOScientific:2007fwp}, and can be simple to implement, we find that if applied here, unless they can be maintained at ultralow temperatures, the associated thermal noise will prevent the quantum noise floor from being reached. This is discussed in detail in Sec.~\ref{ssec:th_noises}. We will therefore assume that the sensor is levitated throughout the rest of this work. 

Assuming that the vertical beam is at normal incidence, and has a width (slightly) exceeding that of the sensor, the force on the particle in the vertical direction (to first order in the geometric optics approximation) is  $F_z\approx 2P_V\, \hat{\bf{z}}$ \cite{ASHKIN1992569}, where $P_V$ denotes the power of the vertical beam. To balance the gravity of the plate, we require $P_V=mg/2\approx 300$~W, where $g$ is the gravitational acceleration on Earth. Although this power requirement is high, it is still within the range of off-the-shelf lasers available  at this wavelength. 

Further, although the majority of power from the levitating beam is expected to reflect off the reflective coating, some fraction will penetrate it and instead be absorbed. This heats the sensor, resulting in the emission of blackbody radiation some of which will then be absorbed by the cryogenic container walls. Heating of the substrate by absorption of light has been studied in Ref.~\cite{Monteiro:2017cmc}, where it was found that for \textit{low} aspect ratio sensors, there is a maximum mass that can be trapped before the sensor reaches melting point. This upper limit arises because the increase in the cooling of the substrate (which scales with the sensor area) is less efficient than the increase in absorption (which scales with volume) when scaling up the mass, such that heavier sensors reach  higher equilibrium temperatures.\footnote{The presence of impurities which affect $\alpha$ as in~\cite{Monteiro:2017cmc} can lower the melting temperature, significantly reducing the maximum mass that it is possible to levitate.} Given $\alpha=10^{-6}$cm$^{-1}$ for optical grade silica \cite{LIGO_technical_abs}, the maximum mass of low aspect ratio sensors is $\approx0.2~\text{mg}$ (see Sec.~\ref{ssec:th_noises}). Particles of this mass have already been proposed for optical levitation in Refs.~\cite{Guccione_2013,Michimura:2016urt}. We thus adopt a sensor mass of $m=0.2$~mg as a short-term benchmark in the rest of the paper. This is a conservative choice given the higher aspect ratio of our design, resulting in an equilibrium temperature that is well below melting point.\footnote{It is therefore possible to decrease the aspect ratio of our 0.2~mg sensor somewhat without significantly heating the system. However the high aspect ratio deployed here was selected to showcase near-resonance `SQL' noise, as discussed in Sec.~\ref{sec:noises}.}

In the future we envisage levitating heavier objects by further increasing the aspect ratio. This would allow  the upper limit on the sensor mass imposed by the melting temperatures of the coating and substrate to be completely evaded, by guaranteeing that the absorbed and emitted powers scale identically with the sensor mass. This is due to the width of each layer of the coating being smaller than the wavelength, such that the coating behaves as a dielectric emitter whose blackbody radiation scales with volume. In contrast, the substrate emits blackbody radiation from its surface. For sufficiently high aspect ratios however the surface area scales approximately with volume. Therefore by growing  the $\hat{\bf{x}}$ and $\hat{\bf{y}}$ dimensions of the sensor without increasing its height $h$, the increase in both the coating volume and the substrate area would scale identically to the absorbed power, i.e.~with the sensor volume (mass). This would allow arbitrarily heavy masses to be levitated without melting provided the increased power requirement on the vertical beam can be met.  In practice however, we may slightly increase the height to $h\approx 20~\mu$m to allow for easier fabrication and handling of the sensor. Although this may slightly increase the equilibrium temperature of the silica substrate, given that the substrate  temperature when $h = 5~\mu$m is relatively far from its melting point (taken to be 1360~K in vacuum \cite{10.1063/1.555799}), this increase should not prove problematic.
To obtain sufficient optical power to levitate heavier objects, a cavity can be formed between the sensor (which acts as a mirror) and a mirror placed beneath it to enhance the scattering power, subject to optical stability \cite{Michimura:2016urt}. Although a careful study of this scenario is beyond the scope of this work, we estimate that the power required to levitate $1~\text{g}$ is $\approx 1~\text{MW}$, which is within reach for a laser power of $100$~W and a cavity power enhancement factor of $\mathcal{O}(10^4)$. Similar intra-cavity powers are already in use in fundamental physics experiments \cite{Capote_2025}. We note that this power requirement is also within the damage/warping thresholds of our reflective coating, which are typically $\approx 10~\text{MW/mm}^2$ \cite{10.1063/1.4817311}. 
In light of this we set a reasonable long-term goal of 
levitating a sensor of mass $m=0.8$~g with dimensions $(100~\text{mm} \times 100~\text{mm} \times 20~\mu\text{m})$.

\section{Noises\label{sec:noises}}
The main noises impacting optomechanical systems can be broadly classified into quantum and classical noises. The former arise, either directly or indirectly, due to the scattering of optical photons on the sensor. Although fundamental to the system, these noises can be optimised for a given experiment and target DM signal via the choice of experimental parameters. Meanwhile, classical noises relate to motion of the sensor due to the presence of thermal baths with which it is in contact, as well as external noises such as seismic noise. Such noises are not fundamental but are intrinsic to any realistic environment. Their control and mitigation is  key to reach the design goal of quantum noise-limited operation. In the rest of the section we specify the noises for a 0.2~mg sensor, unless explicitly stated otherwise.

\subsection{Quantum Noises\label{ssec:quantum_noises}}

Quantum noises arise in optomechanical systems due to white noise fluctuations of the amplitude and phase of the optical field (i.e.~photons) scattering off the sensor. Although increasing the number of photons (i.e.~the laser power) improves the accuracy of the reconstructed sensor position, increased photon scattering also excites motion of the sensor due to unavoidable recoil. The force power spectral density, denoted $S_\text{FF}(\omega)$, resulting from these recoils, is given by
\begin{align}\label{eq:def_backaction_noise_1}
S_\text{FF}^\text{ba}(\omega)=\frac{\Gamma_\text{recoil}}{x_0^2} ~,
\end{align}
where $\Gamma_\text{recoil} =2P_\text{scatt}\omega_L/m\Omega $ is the mechanical excitation rate (see App.~\ref{app:noises_intro} for further details), $P_\text{scatt}$ is the total optical power scattered by the sensor and is related to the power of a single laser $P_L$ by $P_\text{scatt}=2hP_L/\bar{x}$, $\omega_L$ is the frequency of the laser
and $x_0^2=1/(2m\Omega)$ is the zero-point motion of the harmonic oscillator. This noise is known as `backaction' noise (denoted with superscript `ba'), and is also sometimes referred to as `recoil heating' noise. Note that the backaction force power spectrum is independent of the  frequency $\omega$ and scales linearly with the applied laser power.

In contrast the `shot' or `imprecision' noise that arises from the estimation of the sensor position has a position power spectrum which scales inversely with laser power according to
\begin{align}\label{eq:def_shot_noise_1}
    S_{\rm xx}^\text{imp}(\omega)= \frac{x_0^2}{4\,\Gamma_\text{meas}} ~.
\end{align}
Here $\Gamma_\text{meas}$ is the position measurement rate of the readout, which is related to the recoil rate by the detection efficiency $\eta_D\leq 1$ such that $\Gamma_\text{meas}=\eta_D\,\Gamma_\text{recoil}$. Since the majority of photons which are scattered by the reflective coating  are expected to return to the lenses and be read out, we assume ~$\eta_D\approx 1$. Note that the shot-noise position power spectrum is also independent of the frequency $\omega$.

Although $S_{\rm xx}$ corresponds to the experimentally measured observable (position), it is more intuitive to compare the DM induced force to the force noise power spectrum $S_\text{FF}$. This can be straightforwardly obtained from the position power spectrum using the relation $S_\text{\rm xx}(\omega)=\abs{\chi(\omega)}^2 S_\text{FF}(\omega)$, where $\chi(\omega)=m^{-1}(\Omega^2-\omega^2+i\gamma\omega)^{-1}$ is the mechanical susceptibility of the sensor, and  $\gamma$ is the mechanical damping rate, which we assume is much smaller than $\Omega$ (i.e.~see below and Ref.~\cite{Hamaide_2025} for details). 

The total quantum force noise spectrum is the sum of the shot noise and backaction noise spectra, assuming, as we do throughout this work, that their amplitude and phase fluctuations are uncorrelated. It has been shown that (anti-)correlating these (e.g.~by using `squeezed' light) can lead to significant reductions in the total quantum noise \cite{Clerk_2008,Backes_2021,PhysRevX.13.041021,lee2025impulsemeasurementsenhancedsqueezed}. We leave exploration of the potential advantages of such techniques in our setup to future work.

Shot and backaction noise are not specific to optically trapped setups but inherent to \textit{any} general optomechanical system. The minimum total quantum noise that can be achieved with respect to the various degrees of freedom of the system  defines the `Standard Quantum Limit' (SQL). 
Several works have derived expressions for the SQL by minimising the quantum noise with respect to the laser power. They show that, for each frequency of interest, the SQL is reached at the laser power for which the shot and backaction noises are equal. These treatments make the key assumption that the mechanical frequency of the system is independent of the laser power and thus remains fixed upon this optimisation. If the mechanical frequency is itself variable, further enhancements to the sensitivity can be achieved by tuning $\Omega$ to match the target frequency such that the SQL is achieved  `on resonance'. We will henceforth refer to this optimisation procedure - which at present is the only documented systematic strategy for optimising noises in optomechanical systems -  as the `2-stage' approach. 

Whilst appropriate for some types of optomechanical systems (e.g.~suspended cavity mirrors \cite{Carney:2019cio} which use optical fields mainly to read out the mirror's position and velocity), the two-stage treatment cannot be consistently applied to optically trapped devices  as their mechanical frequency depends on the laser power as shown in Eq.~\ref{eq:SHO_frequency}. These parameters therefore cannot  be independently varied and it is in general not possible to consider minimising the noise and tuning the system to resonance as separate procedures. This is also true for cavities whose mechanical frequencies are tuned using the optical spring effect \cite{Aspelmeyer:2013lha}. There has yet to be a  systematic optimisation of the quantum noises for systems with this behaviour in the literature.  We rectify this void here, performing a consistent optimisation of the quantum noises for our system, carefully accounting for these dependencies. Whilst the conclusions we make will be specific to our system, the calculational procedure outlined could be applied to determine the minimum quantum noise for any optically trapped mechanical system. 

Of the three tunable parameters
of our system, $\Omega, \bar{x}$ and $P_L$, only two can be varied independently, with the third then fixed through  Eq.~\ref{eq:SHO_frequency}. For the purpose of deriving a theoretical expression for the SQL, we henceforth work with $\Omega$ and $\bar{x}$, noting that this choice is arbitrary\footnote{In practice it is $\bar{x}$ and $P_L$ which would be tuned in the lab.} and has no bearing on the final result. To optimise the quantum noises we work with their force, as opposed displacement spectra. Since the DM force spectrum is independent of both $\Omega$ and $\bar{x}$, minimising the total quantum force noise spectrum thus directly corresponds to maximising the experimental sensitivity (i.e.~the signal-to-noise ratio).\footnote{Note that the DM displacement spectrum depends on $\Omega$ and would therefore also need to be taken into acount if working instead in terms of $S_{\rm xx}$.  } 

By noting that  
\begin{align}\label{eq:noise_notation}
    \nonumber \frac{\Gamma_\text{recoil}}{x_0^2} &=4\,\omega_LP_\text{scatt} \\ &= 2 \omega_L\bar{x}\,m\,\Omega^2 ~,
\end{align}
where we have used that $P_\text{scatt}=2I_0A(x_r/\bar{x})=\bar{x}\,m\,\Omega^2/2$, the total quantum noise force power spectrum can be expressed in terms of $\Omega$ and $\bar{x}$ according to 
\begin{align}
S_\text{FF}^\text{tot}
    = 2\,\omega_L\bar{x}\,m\,\Omega^2+\frac{m^2[(\Omega^2-\omega^2)^2+\gamma^2\omega^2]}{8\,\omega_L\bar{x}\,m\,\Omega^2} ~.
\end{align}
The SQL is obtained from minimising this expression with respect to $\Omega$ and $\bar{x}$ simultaneously. For a given frequency of interest $\omega$, the minimum noise with respect to $\Omega$ (now with fixed $\bar{x}$) occurs when  \begin{align}\label{eq:min_noise_condition_1}
    (1+16\,\omega_L^2\bar{x}^2)\,\Omega^4=\omega^4 ~. 
\end{align} 
In reaching this result we have made the approximation that $\omega^2 \gg \gamma^2$ which holds over the entire frequency range of interest in our system. 

The minimum quantum noise with respect to $\bar{x}$ (with $\Omega$ fixed) occurs when 
\begin{align}\label{eq:min_noise_condition_2}
    (1+4\,\omega_L\bar{x})\,\Omega^2=\omega^2 ~,
\end{align}
where we have again neglected the term containing $\gamma$ in Eq.~\ref{eq:noise_notation}. Taking the limit $(\frac{2}{5}\omega_L\bar{x})^2\gg 1$ as appropriate for our system (given the range of viable $\bar{x}$), both of these conditions reduce to
\begin{equation}
\label{eq:noisemin}
4\,\omega_L \bar{x}\,\Omega^2 = \omega^2 ~,
\end{equation}
which thus defines the SQL for our system. This expression can be recast as a condition on $\bar{x}$ and the laser power $P_{L}$ as 
\begin{equation}\label{eq:noise_condition_power_laser}
\frac{16\,\omega_L P_L h }{m \bar{x}}=\omega^2~.
\end{equation}

Given that for our 0.2~mg sensor, $\bar{x}$ is constrained to be greater than 5~$\mu$m (the sensor height), it follows that $4\,\bar{x}\,\omega_L\gg 1$. This relation is also true for the $0.8$~g sensor for which the minimum value of $\bar{x}$ is increased to 20~$\mu$m on account of its larger height. It is then immediate from Eq.~\ref{eq:noisemin} that the minimum quantum noise, and thus greatest sensitivity, is not achieved on resonance for either sensor mass.  This result can be intuitively understood as follows: within the physically viable parameter ranges of our system, the backaction noise  \textit{always} dominates over the shot noise when on resonance. To decrease the total quantum noise, the parameters of the system should be varied in such a way to decrease the backaction, in turn raising the shot noise, until the two contributions become equal. The mechanical frequency of the system also decreases as a result of this procedure however, taking the system away from resonance. The fact that SQL does not occur on resonance is not generic to optically trapped devices, but a consequence of the physical restrictions on the parameters of our system. Were it possible to achieve SQL on, or very close to, resonance, there would be a further enhancement to the sensitivity. Nonetheless the above analysis concretely proves that it is better to operate at SQL off resonance as opposed to operating on resonance but above SQL. 
 
When Eq.~\ref{eq:noisemin} is satisfied, the total quantum noise of the system is  
\begin{align}\label{eq:min_noise_SFF}
    S_\text{FF}^\text{tot,min} =4\,\omega_L\bar{x}\,m\,\Omega^2 = m\,\omega^2 ~.  
\end{align} 
This gives $S_\text{FF}^\text{tot,min}= 1.2\times10^{-38}\,(f/4~\text{Hz})^2\,\text{N}^2/\text{Hz}$.
This corresponds to the smallest quantum noise that can be attained in our system for each target frequency and is what will be deployed in the sensitivity forecasts presented in Sec.~\ref{sec:results}. 

We note that, as for the two-stage approach, the backaction and shot-noise contributions to the total quantum noise at SQL are equal. In addition, we find that the value of the quantum noise at SQL, as given by the second equality of Eq.~\ref{eq:min_noise_SFF}, actually matches the off-resonance SQL noise  that would be expected if the  mechanical frequency and power could be varied independently  (i.e.~performing just the first step of the two stage procedure) \cite{Magrini:2020agy,Rossi2018}. This follows from the fact that, for any choice of $\omega$, SQL in our system occurs in the regime $(\omega-\Omega)/\gamma\gg 1$ where $\gamma$ corresponds to the width of the resonant feature in the $S_\text{FF}$, such that it does not affect the noise.
Our system is thus sufficiently far from resonance that the SQL noise is independent of $\Omega$.

We note that the SQL condition stated in Eq.~\ref{eq:noisemin} leaves a degeneracy between $\Omega$ and $\bar{x}$ such that there is not a unique mechanical frequency (or value of $\bar{x}$), at which SQL for each target frequency can be attained. Whilst all choices of ($\Omega, \bar{x}$) which satisfy Eq.~\ref{eq:noisemin} (accounting for the constraints on $\bar{x}$) yield an equivalent sensitivity, they differ in their ease of implementation and practical considerations. From Eq.~\ref{eq:noise_condition_power_laser} it is immediate that minimizing $\bar{x}$  minimizes the horizontal beam power required to reach the SQL. Minimizing laser power is particularly important both when trapping larger masses and at higher frequencies, where these requirements become sizeable.  Recalling that $\bar{x}\gtrsim h/\text{NA}\approx h$, we approximate $\bar{x} \approx h$ (such that $P_\text{scatt}=2P_L$) throughout the rest of this work. In this case, the power requirement on each of the beams directed along the  $\hat{\bf x}$-axis to reach the SQL is $P_L=250~\text{W}~(m/0.2~\text{mg})(f/1~\text{kHz})^2$.

Whilst the restriction on $\bar{x} \approx h \gg x_r$ means that the minimum noise in our system necessarily occurs off resonance, were it possible to achieve SQL on resonance the sensitivity would be further enhanced. In the future, the ideal optomechanical sensor would therefore not only have a tunable mechanical frequency such that a resonant scan over frequency space could be performed, but crucially allow for the SQL to be attained at each resonant frequency. Let us explore how close to resonance the SQL can be obtained in our system. Setting $ \bar{x}  \approx h=5~\mu$m in Eq.~\ref{eq:noisemin} yields that at the SQL, $\Omega \approx 0.03~\omega$. Our system is thus capable of achieving the SQL at each frequency substantially closer to resonance than  other optomechanical devices e.g.~suspended cavities exploiting the optical spring effect where typically $\omega/\Omega\sim 10^2$ \cite{Matsumoto:2018via,Carney:2019cio} (see Ref.~\cite{Hamaide_2025} for further discussion). This is a direct result of our intentionally small choice of $h$ which enables small values of $\bar{x}$. Although lowering the aspect ratio of our sensor (necessarily increasing the value of $\bar{x}$ to follow $h$)  would not change the value of the SQL noise in our setup in Eq.~\ref{eq:min_noise_SFF},   it would alter the value of $\Omega$ required to reach the SQL, resulting in the system being further from resonance. Whilst our system remains in a regime in which the `closeness' to resonance does not impact the sensitivity, our design nevertheless marks a crucial step towards the development of a fully tunable system  capable of leveraging the additional sensitivity enhancements that come with operating at SQL on, or close to, resonance. Our system would begin to see such benefits were it possible to lower $\bar{x}$ to the point where $2\,\omega_L\bar{x}\sim\gamma/\omega$, assuming $\omega/\gamma\approx Q>1$. Achieving this would require improved control over the optical field. This is because, while $h$ could in principle be made smaller, $\bar{x}$ remains constrained from below by the Rayleigh range $x_r$ which cannot be made smaller than the optical wavelength $\sim 1~\mu\text{m}$, since the focused optical beam's width cannot be less than the optical wavelength. We note this constraint also limits the sensitivity of levitated nanoparticles, which scatter light in the Rayleigh regime (see Refs.~\cite{Millen_2020,Hamaide_2025}).
 
As seen in Eq.~\ref{eq:min_noise_SFF} the minimum quantum noise in our system scales as $S_\text{FF}^\text{tot, min}\propto m$. This follows from the fact that in the geometric optics regime the scattered power $P_\text{scatt}$ scales with the cross-sectional area $A$ of the sensor facing the laser. As discussed in Sec.~\ref{sec:dm_couplings}, the DM forces of interest increase linearly with the sensor's mass, such that signal power spectrum $S_\text{FF}^{\rm DM}\propto m^2$. As a result, the signal-to-noise ratio scales as SNR~$\propto \sqrt{m}$, motivating trapping larger masses. This beneficial scaling does not extend over all sensor masses however. Specifically, if the characteristic length scale of the sensor becomes smaller than the wavelength of the trapping optical field, the scattering cross-section $\sigma$ entering $P_{\rm scatt}$, which should now be computed in the Rayleigh regime, depends more strongly on the particle's dimension $R$ since the light interacts with the whole volume of the particle. This is the case for nanoparticles (i.e.~with $R \ll \lambda$) which have historically benefited from interest due to the variety of their applications (see e.g.~\cite{gosling2024levitodynamicspectroscopysinglenanoparticle,tseng2025searchdarkmatterscattering,pontin2022directionalforcesensinglevitated,doi:10.1021/nn901889v,rossi2024quantumdelocalizationlevitatednanoparticle,Hempston_2017,Vinante_2019,Moore_2021}). In this instance the scattered power and quantum noises\footnote{Note that scaling the mass of nanoparticles  does not affect their mechanical frequency $\Omega$ \cite{Millen_2020,Hamaide_2025}.} scale identically with mass to the dark matter signal. It thus follows that upscaling the sensor mass only proves  beneficial to the sensitivity once the geometric optics regime has been reached. 

Finally, we note that the lower surface of the sensor off which the vertical beam reflects, needs to be both smooth and flat. Reflection on uneven sections of the surface will impart horizontal momentum to the reflected photons, leading to recoil of the mirror and thus backaction noise in the sensing direction. Provided these photons are not measured (i.e~they do not enter the detector), they do not contribute to the shot noise. Assuming similar variance in the surface angle as LIGO test masses \cite{Yamamoto:2009,Billingsey:2017}, we expect $9$~mW of vertical power to leak horizontally, corresponding to an additional backaction of
\begin{align}\label{eq:noise_vertical_backaction}
    S_\text{FF}^\text{ba, leak} = 7.3 \times 10^{-39} \text{ N}^2\text{/Hz}~.
\end{align}
This remains subdominant to $S_\text{FF}^\text{tot, min}$ above $f=4$~Hz.

\subsection{Classical Noises\label{ssec:th_noises}}

\begin{figure*}[t!]
    \centering
    \includegraphics[width=1.98\columnwidth]{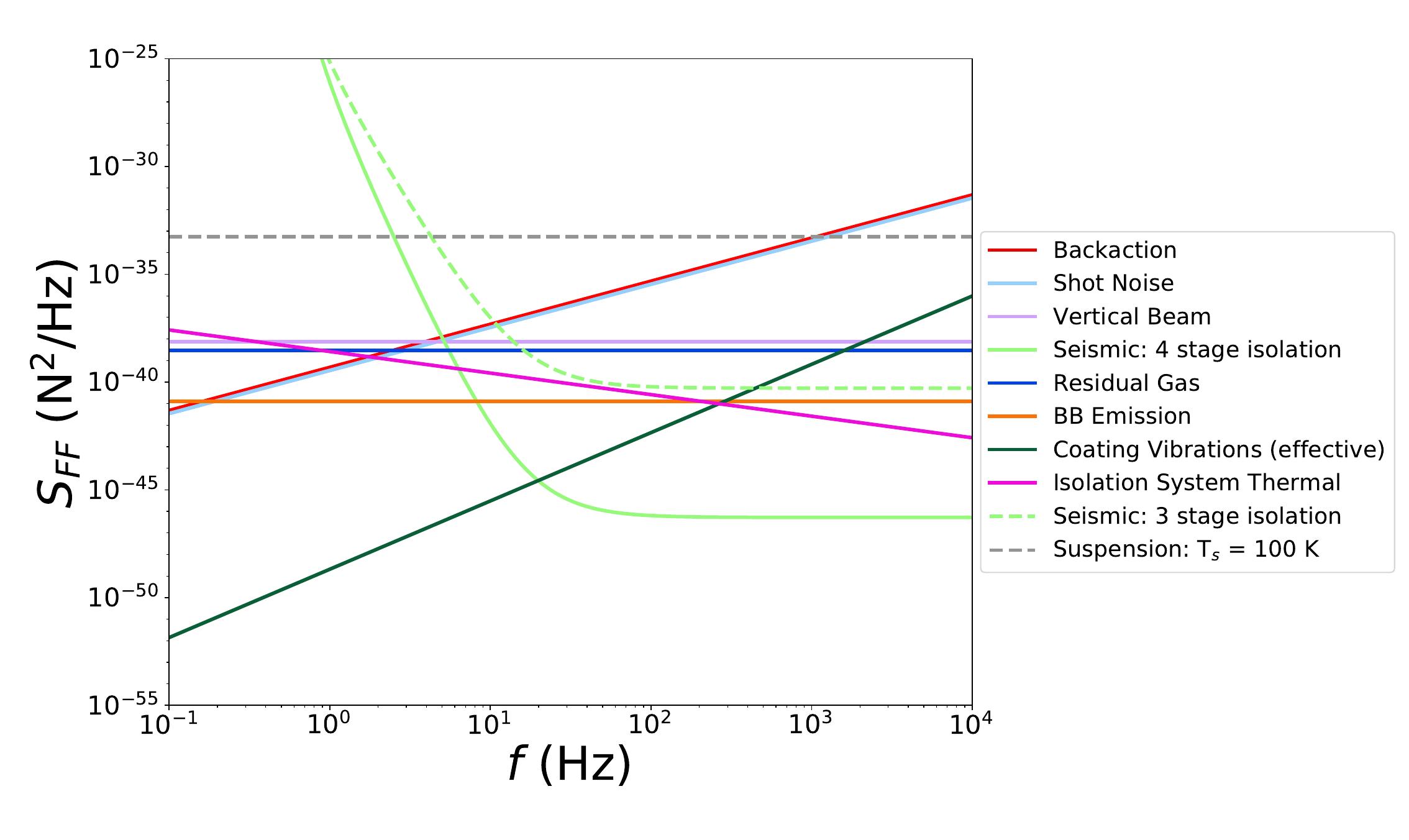}
    \caption{Comparison of both classical and quantum force noise spectra present in our setup for $m=0.2$~mg. Classical noise spectra remain identical for all trapping frequencies, while backaction (red) and shot noise (light blue) are plotted according to Eq.~\ref{eq:min_noise_SFF}, i.e.~are locally correct for any given $\text{f}=\omega_{\rm DM}/(2\pi)$. We ignore the effect of possible additional heating due to the trapping beams above 5~kHz, as quantum noises dominate in this region. The coating's vibrational noise (dark green) and blackbody radiation (orange) are subdominant to quantum noises at all frequencies, while residual gas (dark blue) is subdominant above $5$~Hz by design. Below this point, the seismic noise (solid green) dominates, while vertical beam-induced horizontal recoil noise (lilac) is coincidentally nearly identically to quantum noises at the quantum/seismic transition. For comparison, we show in dashed green the seismic noise for the 3-stage seismic isolation system (dashed green) used in Ref.~\cite{Deshpande:2024bul}. We finally show that levitation is preferable to suspension of our sensor since its noise (dashed, assuming a typical damping rate and a string temperature of $T_s \approx$ 100~K) is significantly larger. See Sec.~\ref{sec:noises} for further discussion.}
    \label{fig:noises}
\end{figure*}

Having optimised the quantum noises, we now turn to classical noises and the conditions necessary to ensure that they remain subdominant in the proposed experiment. We consider classical noises  originating from external thermal baths with temperatures $T_i>0$, seismic motion of the external environment and material heating through the absorption of optical photons. We treat each noise separately, beginning with thermal noises from emitted blackbody radiation and residual gas in the vacuum chamber. For each thermal bath, we may describe the spectrum of motion as Brownian motion according to the fluctuation-dissipation theorem which states that \begin{equation}\label{eq:fluctuation_dissipation_th}
   S_{\rm xx}^{T_i}(\omega)=\frac{2k_BT_i}{\omega}\abs{\text{Im}(\chi(\omega))} ~,
\end{equation}
where $\abs{\text{Im}(\chi(\omega))}\approx \gamma_i/(m\omega^3)$ given that $\omega^2\gg\Omega^2\gg\gamma_i\omega$ for our system. Here $\gamma_i$ refers to the specific damping rate from the thermal bath in question.
Throughout this section, we will initially obtain  $S_\text{xx}$, which is then  converted to $S_\text{FF}$ using the mechanical susceptibility of the centre of mass motion of the sensor $\abs{\chi(\omega)}^2$, unless stated otherwise.

\subsubsection{Blackbody Radiation}
Substantial heating of the system may occur through the absorption of optical photons. This is expected to mainly take place in the coating (as opposed to the substrate). The absorbed power is given by
\begin{equation}
\label{eq:Pabs}
P_\text{abs}\approx \alpha P_L R_i~,
\end{equation} 
where $R_i$ is the length of the coating along the beam axis and $\alpha \approx 6\times 10^{-3}~\text{cm}^{-1}$ is the absorption coefficient of the coating~\cite{Steinlechner:22,Steinlechner_2018,Chalermsongsak_2016}. 
As a Bragg mirror, decreasing amounts of light are reflected off each Si/Ta interface in the coating, such that most of the light only penetrates the first few layers. This decreases the effective absorption length $R_i$, in turn reducing the total absorption. Since the power of the vertical beam exceeds that of the horizontal trapping beams up to $\Omega=1$~kHz, this dominates $P_\text{abs}$, which is therefore independent of frequency up to $\Omega\approx 1$~kHz, i.e.~in almost the entire frequency space of interest. Using this in Eq.~\ref{eq:Pabs}, we estimate the absorbed power to be $P_\text{abs}=0.1$~mW.

Given the thickness of each layer of coating is $\mathcal{O}(0.1~\lambda_L)$, the coating will cool by emitting blackbody radiation orthogonally to its surface as a dielectric (i.e.~in the Rayleigh emission regime), according to \cite{Chang_2009}

\begin{align}\label{eq:bb_radiation}
    P_\text{BB}\approx\frac{72\,\text{Im}(\alpha_p)\,V}{\pi^2} (k_BT)^5 ~,
\end{align}
where $V = 3.5~\rm{mm} \times 3.5~\rm{mm} \times 0.47~\mu\rm{m} $ is the volume of the coating's outer layer, $\text{Im}(\alpha_p)$ is the imaginary part of the layer's polarizability (assumed to be approximately constant in $\omega$), and $T$ is the coating temperature. The equilibrium temperature of the sensor is that which satisfies $P_\text{abs}=P_\text{BB}$. 
Conservatively assuming blackbody radiation is emitted only by the outermost layer of coating, and taking the value of $P_{\rm abs}$ computed from Eq.~\ref{eq:Pabs} (where absorption occurs over the entire coating), we find $T\approx 160$~K. 
We note that blackbody radiation occurs in all directions, such that some of it will not be collected by the optical lenses, but instead be absorbed by the surrounding walls of the cryogenic box. However, the resulting heating can be more than compensated for by the cooling capabilities of a 4~K closed-cycle liquid helium cryostat employing a pulse tube refrigerator used in low-vibration setups such as that described in Ref.~\cite{Deshpande:2024bul}, thus avoiding a significant rise in the temperature of the system. 

The emission of blackbody photons will lead to recoil of the sensor. The power spectrum of the induced motion can be obtained  using the fluctuation-dissipation theorem (Eq.~\ref{eq:fluctuation_dissipation_th}) analogously to thermal baths, with the damping rate \cite{rahman2021realisingeinsteinsmirroroptomechanical}
\begin{align}
    \gamma_\text{BB}=\frac{2\,P_\text{BB}}{m} ~.
\end{align}
Using this we obtain $S_\text{FF}^\text{BB}=1.3\times 10^{-41}\text{N}^2/\text{Hz}$ for $T=160$~K.

We note the silica substrate will also absorb light, primarily from the levitating beam. However, not only does silica have ultralow absorption rates ($\alpha\approx 10^{-6}$~cm$^{-1}$) but it will reflect some of the 1064~nm laser light at its surface \cite{Billingsey:2017}. In the case of the 0.2~mg sensor, the substrate is not expected to heat as much as the coating in the frequency range of interest. Further, for the $m\sim0.8$~g mass sensor, the temperatures of both the substrate and coating are expected to remain similar to those of the $m\sim0.2$~mg mass sensor, as a result of the increased aspect ratio as discussed in Sec.~\ref{sec:setup}.

\subsubsection{Residual Gas}
Collisions of residual gas particles present in the vacuum chamber against the sides the sensor present a further source of thermal noise. We expect the majority of gas scattering events to be diffuse, i.e.~with isotropically distributed outgoing angles, rather than specular, i.e.~equal ingoing and outgoing angles~\cite{Martinetz_2018}. We conservatively take the accommodation coefficient of our polished surface, i.e.~the fraction of scattering events that are diffuse, to be $\alpha_c\approx 0.8$ \cite{Bayer_Buhr_2022}. Given that the surface area of the top and bottom of the sensor is much larger than the area of the sides, the dominant source of gas force noise in the $\hat{\bf x}$-direction comes from diffuse gas scattering on the top and bottom surfaces. Were 100\% of the gas collisions with these surfaces specular, the power spectrum of the resulting displacements (which would be entirely in the $\hat{\bf z}$  direction i.e.~yielding an $S_{\rm zz}$), would follow from the fluctuation dissipation theorem (Eq.~\ref{eq:fluctuation_dissipation_th}), taking the (specular) damping rate on the sensor's motion to be that of an ideal gas in the Knudsen regime \cite{Jain:2016,Millen_2020}
\begin{align}
    \gamma_\text{gas}\approx \frac{PA}{mv_\text{gas}} ~. 
\end{align}
Here $v_\text{gas}=\sqrt{3k_BT/m_a}$, where $m_a$ is the mass of a (nitrogen) gas molecule, $P = 10^{-12}$~mBar corresponds to the residual gas pressure in the vacuum chamber and $A$ is the area of the top/bottom surface of the sensor (i.e.~$A=3.5~\text{mm} \times 3.5~\text{mm}$). We assume that the gas temperature is $\approx 100 $~K, a reasonable middle ground between the temperature of the sensor and the liquid helium cryogenic container of $\sim$ 4~K, with which the gas is in contact.

In diffuse scattering the momentum transfer to the surface is instead isotropically distributed, thus giving rise to force noise in the $\hat{\bf x}$ direction. To obtain an estimate for the power spectrum of the displacements in the $\hat{\bf x}$ direction we may therefore multiply $S_\text{\rm zz}$ by the square of $\alpha_c$ times ~$2/\pi$ (the fraction of the isotropic momentum transfers which are in the $\hat{\bf{x}}$ direction) i.e.~$S_{\rm xx} = (2\alpha_c/\pi)^2 S_{\rm zz}$.  On conversion to an $S_{\rm FF}$ this yields $S_\text{FF}=2.9\times 10^{-39}~\text{N}^2/\text{Hz}$, which is subdominant to the quantum and seismic noises of the system across all frequencies of interest.

\subsubsection{Seismic Noise}
Another important source of noise is seismic noise. This couples directly to the lab, and hence the plate on which the optical trap lies. This leads to a relative percieved motion of the sensor relative to the plate (reference frame). Seismic noise is especially problematic at low frequencies where measured unmitigated seismic noise displacements are large and follow a power law scaling $S_{\rm xx}(\omega)\sim 1/\omega^2$ \cite{galaxies10010020}. In our setup seismic noise is mitigated through  use of a 4-stage seismic damping system which supports the optical components of our horizontal and vertical beams (see Sec.~\ref{sec:setup} for details), similar to the 3-stage system used in Ref.~\cite{Deshpande:2024bul}. This system strongly reduces seismic noise from the external environment, particularly at frequencies above $\sim$ 0.5~Hz where the resonant (pendulum) modes of the suspended plates (i.e.~stages of the isolation system) are assumed to lie. At these frequencies there is a strong power law suppression of the seismic motion with the inferred force power spectrum  scaling as $S_\text{FF}^\text{inf}(\omega)\sim 1/\omega^{4n}$, where $n$ denotes the total number of stages (i.e.~4 for our setup). To predict the seismic noise force power spectrum in our system we multiply raw (unmitigated) seismic displacment spectral data, which was measured above ground at Northwestern University and fitted to be $S_{\rm xx}(f)\approx10^{-10}~(\text{m}/\sqrt{\text{Hz}})~(100~\text{Hz}/f)^2$, by the transfer function of the isolation system $\chi^\text{s}(f)$. Assuming $n$ geometric anti-springs (GAS) each with damping rate $\approx 10^{-3}$ we may use the following fitted transfer function for our system (based on the system in Ref.~\cite{Deshpande:2024bul})
\begin{align}
    \chi^\text{s}(f)=\abs{\frac{0.25 + 10^{-3}f^2 }{ 0.25 - f^2}}^{2n}~,
\end{align}
to yield $S_\text{FF}^\text{inf}=9.9\times 10^{-43}~\text{N}^2/\text{Hz}~(f/10~\text{Hz})^{-16}$ for $m=0.2$~mg at low frequencies ($f\lesssim 10$~Hz), above which it converges to $S_\text{FF}^\text{inf}\approx 10^{-46}$. This may be reduced further by performing the experiment in an underground facilty where raw seismic noise is expected to be 2-3 orders of magnitude lower (depending on the location) compared to at the Earth's surface~\cite{galaxies10010020}. 

Further seismic noise mitigation may also be achieved by making differential measurements between a pair of sensors with different charges under the DM force as discussed in Ref.~\cite{Carney:2019cio}. This would allow for the rejection of common seismic modes. Whilst the system would remain impacted by seismic gradient noise, this would only represent $\mathcal{O}(0.01\%)$ of the original seismic noise at $\mathcal{O}(1~\text{Hz})$ for a sensor separation of $\mathcal{O}(1~\text{cm})$. We leave a more detailed discussion of the practicalities of using this technique and its effect on our setup's noise spectrum to future work.

\subsubsection{Isolation system thermal noise}
The seismic isolation system is expected to introduce thermal noise from its physical contact with the optical setup. The thermal displacement noise of the seismic isolation system's final stage, $S_{\rm xx}^\text{susp}$, can be obtained with the fluctuation-dissipation theorem using the platform's mechanical susceptibility. In this work we adopt the same mechanical susceptibility as Ref.~\cite{Geraci_2019}, which includes both coupled vibrational and centre of mass motion, correcting  for the different mass of our plate (which we assume to be 1~kg). We note that the damping rate that enters this susceptibility is similar to that of the suspension mechanism in Ref.~\cite{Matsumoto:2018via} and thus deemed to be an appropriate benchmark. We assume that the temperature of the plate\footnote{This contrasts to Ref.~\cite{Geraci_2019} where room temeprature ($\sim$ 300~K) was assumed.} is $T=100$~K i.e.~similar to the residual gas temperature. This leads to $S_{\rm xx}^\text{susp}\propto \omega^{-5}$ as a result of the coupled vibrational and centre of mass motion of the optical setup. As with seismic noise, the motion of the plate  leads to a perceived motion of our sensor in the plate's frame of reference. To convert this perceived motion to an estimated $S_\text{FF}$ on the sensor we use the sensor's mechanical susceptibility which notably leads to $S_\text{FF}^\text{susp}\propto m^2$. This way, we obtain $S_\text{FF}^\text{susp}=4\times 10^{-38}/\omega~\,(\text{N}^2/\text{Hz})$, which is subdominant to the combined quantum and seismic noises in the entire frequency range of interest.

\subsubsection{Newtonian Noise}
A further non-thermal source of noise is Newtonian gravitational noise. Time-varying Newtonian forces in the sensing direction from both seismic and anthropogenic sources are expected to accelerate both the final stage of the seismic isolation system as well as the sensor. 
The force on the sensor is inferred from the relative acceleration of the sensor compared to the isolation system's plate. Given that both are co-located, were they ideal free particles, they would experience identical accelerations from Newtownian gravtiational forces. In practice, since  both the sensor and the plate are trapped, a small difference in acceleration is expected due to the difference in their mechanical susceptibilities. We estimate the difference in mechanical susceptibility to be of order 0.5\% in the frequency range of interest. Assuming that the Newtonian gravitational fields acting on our system are similar to those at LIGO \cite{galaxies10010020}, the induced Newtonian force noise in our system is expected to be subdominant to seismic and quantum noises across the entire frequency range of interest.

\subsubsection{Coating vibrations}
So far, we have dealt with motion from the sensor's centre of mass (CoM), however thermal baths can also excite vibrational modes of the sensor, deforming the surfaces facing the $\pm\hat{\bf{x}}$ lasers. This leads to noise in the sensor's reconstructed CoM position. 
The mechanical susceptibilities of the substrate and coating harmonic oscillator modes can be modelled as in Refs.~\cite{Gras_2018,Levin_1998,Harry_2002}, where they are shown to be in good agreement with data taken at LIGO (which uses a very similar coating and substrate to our setup).
The total vibrational noise spectrum is then given by
\begin{align}\label{eq:noise_thermal_total}
    S_\text{\rm xx}^\text{vib}(\omega)= \frac{2k_BT}{\omega} 
    \sum_\text{v}\abs{\text{Im}(\chi_\text{v}(\Omega_v,\omega))} ~,
\end{align}
where $T$ is the temperature of the sensor and the sum is taken over all vibrational modes which affect the surface's measured position.
Note that due to the higher aspect ratio of the coating, the contribution from the coating modes dominates over that from the substrate modes. Further, as $S_{\rm xx}^\text{vib}$ contains modes with different $(\gamma_i,m_i,\Omega_i)$ (where $m_i$ are the reduced masses of vibrational modes), it can only be converted to an \textit{effective} force noise  which we define as $S_\text{FF}^\text{eff}=\abs{\chi(\omega)}^{-2}S_{\rm xx}$ \cite{Manley:2020mjq}.  Here $\chi(\omega)$ represents the CoM susceptibility (which is approximately independent of $\gamma_i$). The quantity $\text{Im}(\chi_\text{v})$ is typically independent of $\omega$  since we are in the regime $\omega\ll\Omega_v$ \cite{Michimura:2016urt}. As such $S_{\rm xx}^\text{vib}\sim 1/\omega$, leading an effective $S_\text{FF}^\text{eff}\sim\omega^3$. This contrasts to the CoM thermal force noises of thermal baths $i$ given by $S_\text{FF}^{T_i}=2mk_BT_i\gamma_i$  which are spectrally flat. Further note that although we account for thermal vibrational noise, we neglect vibrations induced by DM since the wavelength of the DM field greatly exceeds the sensor dimensions. 

We note optical fluctuations which scatter off vibrational modes from flexing and torsion of the plate around both of the horizontal directions ($\hat{\bf x}$ and $\hat{\bf y}$) may also contribute to backaction noise due to the angle at which photons would recoil. Although this is expected to average out in the (comparatively low) frequency range of interest for our 0.2~mg sensor given the typically high frequency of these vibrations, it may be necessary to damp this motion for a 0.8~g sensor where the  flexing and torsion frequencies may be lower (see Sec.~\ref{ssec:trap_stability}). To check this, we may estimate the frequency of flexural modes, which generally scale as $h/L^2$, given the dimensions of the sensor and the elastic modulus of the substrate and coatings. Although a comprehensive mode analysis is beyond the scope of this paper as it may vary with particular reflection coating parameters, for $h \approx 5~\mu$m, we expect the lowest frequency modes of our 0.2~mg sensor to be at approximately 10~kHz, thus exceeding the range of frequencies targeted in this work. 

\subsubsection{Frequency and Intensity noise}
We comment on the presence of classical intensity and frequency noises in the optical field. These act as the classical analogs to backaction and shot noise, and have been studied in Ref.~\cite{Michimura:2016urt} where is was found that they are typically subdominant to quantum noises. We thus expect these to be insignificant in our system but note that if necessary, they could be further suppressed through the use of feedback loops and cavities at the laser input stage of the optical field, i.e.~outside the vacuum chamber \cite{Michimura:2016urt}.

\subsubsection{Suspension Noise}
We finally discuss the choice to levitate our sensor rather than suspend it, e.g.~from a string. In the absence of additional active cooling, we assume the string would be at a temperature $T_s \approx 100$~K. Provided the string length is $\geq\mathcal{O}(10~\text{cm})$, the natural frequency of the suspension system will be below the frequency range of interest for DM searches. This ensures that the string would not contribute significantly to the sensor's spring constant (when added in parallel) such that its use would not impede our ability to tune $\Omega$. However, since the suspension mechanism is in direct contact with the sensor, it constitutes an additional thermal bath and thus provides an additional source of thermal noise. Using the fluctuation-dissipation theorem i.e.~$S_\text{\rm xx}^\text{susp}=2k_BT\gamma/(m\omega^4)$, and assuming a damping rate of $\gamma\approx 10^{-6}$~Hz as measured in (high-Q) suspension systems \cite{Matsumoto:2018via,Geraci_2019},
we obtain $S_\text{FF}^\text{susp}=1.5\times 10^{-34}~\,(\text{N}^2/\text{Hz})$. 
This dominates over the quantum noises for the majority of the frequency range of interest. Although suspension is in principle easy to implement, given that the noise spectrum scales linearly with the temperature, in order for the suspension noise to be subdominant to other noises sources, especially at low frequencies,  ultralow temperatures of around 10~mK would be required. Although a full study of a setup using suspension is beyond the scope of this work, we emphasise that achieving these temperatures would likely require additional cooling with a dilution fridge. It would also be necessary to sufficiently shield  the suspension system from blackbody radiation emitted by the sensor coating which, using an analagous analysis as for the levitated case (see Eq.~\ref{eq:bb_radiation}), would be expected to heat to 34~K at 6~Hz due to the absorption of light from the horizontal beams only (given the absence of the vertical beam). Given that the sensor would be in direct physical contact with the suspension system, we thus believe maintaining the suspension system at $T=10$~mK would present a significant challenge. Therefore, levitating the sensor proves to be beneficial to the performance of our setup as the noise from the levitating beam is already expected to be subdominant to the total quantum noise and seismic noise at all frequencies even with a higher sensor temperature $T=160$~K.

\subsection{Discussion}
In Fig.~\ref{fig:noises} we  plot the $S_{\rm FF}$ from blackbody radiation, residual gas and  
vertical beam backaction, as well as seismic noise, thermal noise from the seismic isolation system,
effective coating vibrational noise and the optimised quantum noises (backaction and shot noise) as discussed above, for a $m=0.2$~mg sensor. To highlight the reasoning behind our decision to levitate the sensor rather than suspend it, we plot the thermal noise that would be expected to arise from  suspension, assuming a string temperature of $T_s = 100~$K. We emphasise that at each frequency of interest,  $\Omega$ and $\bar{x}$ have been selected to optimise the quantum noises at that frequency. In practice we set $\bar{x} \approx h$ and then fix $\Omega$ using Eq.~\ref{eq:noisemin}. As we have seen, this means that we are necessarily in the regime  $\Omega \ll \omega$. We use the same values of  
($\Omega,\bar{x}$) in plotting the classical noises. This is thus tantamount to taking them in the limit $\omega\gg\Omega$ at every frequency.  The figure should be interpreted as the local noise power spectra achieved when searching for a DM signal at a given $\omega=\omega_\text{DM}$, as opposed to the spectra attained with any fixed experimental setup. 

As seen in Fig.~\ref{fig:noises}, seismic noise becomes dominant over the quantum noises below $f\approx 5$~Hz. For comparison, in Fig.~\ref{fig:noises} we show the performance of the 3-stage seismic isolation system as used in Ref.~\cite{Deshpande:2024bul} to motivate our choice to add an additional fourth stage. The dominance of seismic noise at low frequencies motivates future exploration of the use of differential measurements between a pair of sensors with different neutron-to-nucleon ratios to suppress common seismic modes and thus recover some of this parameter space.  We note that by setting the residual gas pressure to $P=10^{-12}$~mBar, the thermal  gas noise is subdominant across the entire frequency range of interest. Vibrational noise (which we denote `Coating Vibrations (effective)') is also subleading to quantum noise in the frequency range of interest. It should be noted that at very high frequencies ($f>1$~kHz), the absorption heating will be dominated by the power in the $\bf{\hat{x}}$-direction beams (instead of the vertical beam), such that the sensor temperature will reach melting temperature at $f\gtrsim 200$~kHz,
independently of its mass (since scaling the mass will not affect the sensor temperature, as discussed in Sec.~\ref{sec:setup}). Although the sensor is expected to heat to temperatures above 160~K above 1~kHz, we do not show this effect in Fig.~\ref{fig:noises}. This is because the increased noise from blackbody radiation, vibrations, and from additional heating of the residual gas and the seismic isolation system will have no impact on the sensitivity of our setup as quantum noises dominate other noises by up to ten orders of magnitude at 1~kHz. Therefore pressure and temperature requirements at frequencies above $\approx 5$~Hz are effectively eased compared to those near 5~Hz, given a fixed sensor mass. 

Compared to the 0.2~mg sensor, some further considerations to suppress classical noises might be necessary when using a 0.8~g sensor. Both the quantum noises and the horizontal backaction induced by the vertical beam scale linearly with the sensor mass. Given that the high aspect ratio of our sensor geometry means that $A\sim m$ when scaling the mass, the residual gas noise will also scale linearly with $m$. As such the frequency at which the residual gas and vertical beam noises becomes dominant over the quantum noises will be the same as for the $m=0.2$~mg (see Fig.~\ref{fig:noises}). In contrast, thermal noise from the seismic isolation system, the next subdominant source of noise, scales as $m^2/m_\text{plate}$, where $m_\text{plate}$ is the mass of the seismic isolation system's final stage plate, which we assume will likely increase to accomodate a larger sensor. The plate may be further supplemented with additional thickness to increase its mass in order to leave the thermal noise of the seismic isolation system (approximately) unchanged. However this would not improve the seismic $S_\text{FF}$ noise which goes as $m^2$ and is independent of the plate's mass. Suppression of the seismic noise could occur if differential measurements with a second optical sensor (on the same seismic isolation plate) were performed, which would allow for common mode suppression. Further suppression could also be achieved by using a 5-stage seismic isolation system, the detailed study of which we leave for future work. If we assume either of these changes (or a combination thereof) significantly counteracts the increase in noise perceived for the 0.8~g mass sensor, 
the increased gas and vertical beam backaction noises would dominate over all noises over a range of low frequencies, which in the case of a $m=0.8$~g sensor corresponds to $\approx$~3-5~Hz. Thus in this region, reducing the residual gas pressure to at least $P=10^{-13}$~mBar would yield further sensitivty gains. To achieve lower gas pressures, increased cooling resources (e.g.~to reach $T\sim 10$~mK) may be necessary, along with efficient shielding/cooling from the effects of the sensor's blackbody radiation. Meanwhile lowering the backaction from the vertical beam would necessitate more advanced techniques in sensor fabrication and polishing to produce flatter surfaces (i.e.~with a variance in the surface angle smaller by a factor of up to $10^{-3}$). A more in-depth study of these practicalities is left for future work.

Finally, we note that suspension would not be a viable alternative to improving sensor fabrication at larger masses either, as the associated thermal noise also scales linearly with mass. Therefore, irrespective of the sensor mass, suspension is only useful if the temperature of the string can be brought down to $T_s=10$~mK. Although shielding may help lower the suspension temperature by isolating it from gas heating and/or blackbody radiation, it remains in  unavoidable physical contact with the sensor, whose temperature is expected to be 34~K at 5~Hz and increase further with frequency due to absorption from the horizontal beams, thus making the required string temperature challenging to achieve.

\subsection{Trap stability and sensor control}\label{ssec:trap_stability}

Although this work is restricted to sensing forces in the $\hat{\bf{x}}$-direction, excessive translational or rotational motion in other directions can lead to readout issues, including the deflection of light away from the lenses and thus the photodetectors. This would decrease the detection efficiency $\eta_D$, in turn increasing the perceived shot noise. Further, the escaping photons would be absorbed by the cryogenic vacuum chamber walls risking overwhelming the cooling power of the liquid helium, increasing the temperature of the system and inducing additional thermal noises (e.g.~from residual gas or blackbody radiation). Separately, if motion along these degrees of freedom results in photon recoil in the $\hat{\bf{x}}$-direction, additional backaction force noise will also be generated. This is particularly relevant to rotation around the $\hat{\bf{y}}$-axis i.e.~in the $\hat{\bf\phi}$ direction, where the recoil of photons from the vertical beam would lead to translational forces along the $\hat{\bf{x}}$ axis. To avoid this, and additionally prevent any potential instability (i.e.~non-linearities) of the sensor motion from arising, these degrees of freedom must be controlled. This is done by containing (trapping) each degree of freedom using restoring optical forces to form harmonic oscillator potentials.

First, a cylindrical lens is used to focus the vertical beam in the $\hat{\bf{y}}$-direction. This prevents photons from being given momentum in the $\hat{\bf{x}}$ direction, thus avoiding recoil from absorbed photons along this axis. This lensing also allows rotational motion around the $(\hat{\bf{x}},\hat{\bf{y}})$-axes (i.e.~angular motion in the $(\hat{\bf\theta},\hat{\bf\phi}) $ directions) to be controlled by creating angular harmonic oscillator potentials. Assuming a $\text{NA}\approx 0.8$, both of these angular traps  have a mechanical frequency of $\Omega_q\approx 250$~Hz where $q \in(\theta,\phi)$, and zero-point angles $q_0=\mathcal{O}(10^{-13})$ defined by $q_0=\sqrt{1/I_q\Omega_q}$ where $I_q$ is the moment of inertia. For further details on angular trapping see App.~\ref{app:angular_trap}. Motion in the $\hat{\bf \phi}$ direction generates additional backaction noise in the $\hat{\bf{x}}$ direction. 
To prevent this noise from exceeding the roughness-induced backaction noise from the vertical beam given in Eq.~\ref{eq:noise_vertical_backaction}, which already surpasses the quantum noises at low frequencies, we must ensure that the standard deviation of the angle $\sigma_\phi$ does not exceed the standard deviation of the roughness angle of the mirror surface, taken to be $\sigma_{\phi,\text{rough}}=3\times 10^{-6}$ \cite{Yamamoto:2009,LIGO_technical_abs}. Further, we must also ensure that the standard deviation $\sigma_q$ (where $q \in(\theta,\phi)$) of rotations around the $\hat{\bf{x}}$ and $\hat{\bf{y}}$ axes (i.e.~in the $\hat{\bf{\theta}}$ and $\hat{\bf{\phi}}$ directions) are kept small such that $\sigma_q\ll h/L\approx 10^{-3}~\text{rad}$. Above this value, the horizontal beams would irradiate the bottom and top surfaces of the sensor leading to torques that will destabilize the sensor\footnote{We also note that the alignment (in $\hat{\bf{z}}$) of opposing horizontal beams is required to be within $h=5~\mu$m to avoid instabilities. Note that if necessary, these rotation and alignment constraints could be relaxed by increasing the value of $h$.}. Next, if the motion along the $\hat{\bf z}$-axis leads the sensor to no longer be irradiated by the horizontal beams and assuming $\bar{x}\approx h$ the sensor will escape trap along the $\hat{\bf x}$-direction if vertical motion along $\hat{\bf z}$ has standard deviation $\sigma_z\gtrsim\mathcal{O}(h)$. We must therefore set $\sigma_z\ll h$. We begin by writing 
\begin{align}\label{eq:ang_mech_temp_variance}
    \sigma_q^2=(2n_q+1)q_0^2\approx(2k_BT_q/\Omega_q)q_0^2~,
\end{align}
where $n_q$ is the angular phonon number\footnote{Note this is analogous to motion in the $\hat{\bf{x}}$-direction where $\langle x^2\rangle=(2n_x+1)\, x_0^2$ where $n_x$ is the phonon occupation number of the sensor. For further details, see App.~\ref{app:angular_trap}.} and $T_q$ is defined here as the mechanical temperature. Using Eq.~\ref{eq:ang_mech_temp_variance}, the first condition ($\sigma_\phi<\sigma_{\phi,\text{rough}}$) corresponds to $n_\phi\lesssim 10^{15}$ and $T_\phi\lesssim 10^7$~K, while the second condition leads to even higher bounds on $n_q$ and $T_q$. Finally, the third condition leads to bounds on $n_z$ and $T_z$ which are also well above the gas temperature. Given these upper bounds are well above the expected temperature of both the surrounding thermal bath (around 100~K) and the effective temperature of the optical field (taken to be $T_L\approx \omega_L\approx 10^4$~K, following Refs.~\cite{Millen_2020,Jain:2016}), the latter of which usually constitutes the upper bound on the temperature of any mechanical degree of freedom $q$ in optomechanical systems if no substantial sources of damping (other than that from the optical field's photon bath) are present. If the mechanical temperature is significantly higher than the gas temperature however, the gas will be heated. To avoid this, feedback damping loops need to be applied to each rotational direction to damp their respective motions and thus control their mechanical temperature. For the mechanical temperature of these degrees of freedom to match the residual gas temperature $T=100$~K, we require $n_\phi\approx 10^{10}$. The required damping rate  can be found using the steady-state relationship $n_q=\Gamma_q/\gamma_q$ \cite{Jain:2016}, where the total damping rate is dominated by the feedback damping, i.e.~$\gamma_q\approx\gamma_\text{fb,q}$.


In ultrahigh vacuum, it may be necessary to apply further damping to all the translational, rotational and flexural degrees of freedom,  e.g.~with the use of active feedback loops, to prevent the sensor from escaping the trap. This is particularly important for the degrees of freedom which we have not constrained yet, namely motion along the $(\hat{\bf x},\hat{\bf y})$-axis and rotation  in direction $\hat{\psi}$. Such motion typically arises due to the excitation of phonons by scattering photons, but can also be induced by absorption of circularly polarised light (used for readout) which can lead to torque on the sensor. Similarly to the case of rotation in the $\hat{\bf\phi}$ direction, the necessary damping could be implemented using separate, measurement-based linear feedback cooling loops or by using parametric feedback cooling utilising the light fields used to levitate and trap the sensor. Linear feedback cooling could use separate, weaker optical fields or even electrostatic fields for actuation \cite{Tian2024,Hsu2016,Rossi2018}. To obtain the required damping rates, the mechanical damping rates for each degree of freedom are (conservatively) constrained such that non-linearities of the spectrum of motion remain small relative to the damping rate. Up to an $\mathcal{O}(1)$ numerical factor we may write this constraint as~\cite{Gieseler_2013,Hamaide_2025,Carney:2019cio} 
\begin{align}\label{eq:NL_damping_condition}
    \gamma_{\text{min},q}\gg \sqrt{\frac{ P_\text{scatt}\omega_Lq_0^2}{\bar{q}^2m}}~, 
\end{align} 
where $\bar{q}$ is the scale of the distance/angle at which non-linearities appear in the trapping potential (e.g.~$\bar{x}$ for motion along the $\hat{\bf{x}}$-axis). In the $\hat{\bf x}$-direction, this corresponds to a damping rate of $\gamma_\text{min}\sim 10^{-11}$~Hz at $\Omega_x=5$~Hz. This condition also guarantees that the sensor cannot escape the trap. 

Using that $k_BT_q=n_q\Omega_q=\Omega_q\Gamma_q/\gamma_q$, the condition in Eq.~\ref{eq:NL_damping_condition} yields an upper bound on the mechanical temperature (see App.~\ref{app:angular_trap} for details). Applied to the $\hat{\bf{x}}$-direction this is
\begin{align}\label{eq:freq_condition1}
    T_x\lesssim 1~\text{K}~\left(\frac{\Omega_x}{5~\text{Hz}}\right)^{5/2} ~,
\end{align}
which is lower than the temperature of the surrounding bath (the residual gas at $T\approx 100$~K) for the lowest values of $\Omega_x$ that we intend to use, and which again needs to be implemented using  active feedback damping. This analysis can be repeated for rotations around the $\hat{\bf{z}}$-axis (motion in the $\hat{\bf\psi}$-direction) to yield similar bounds on $T_\psi$ and $\gamma_\psi$ as for the $\hat{\bf{x}}$ direction. For the other remaining degrees of freedom however, namely $(\hat{\bf{y}},\hat{\bf{z}})$, the upper bounds on the mechanical temperatures are typically higher (assuming NA~$\approx 0.8$), due to an increased freedom of motion in these directions. This  allows us to fix $\bar{q}$ to much larger values than we had for $\bar{q}=\bar{x}$  leading to relaxed constraints in  Eq.~\ref{eq:NL_damping_condition} and Eq.~\ref{eq:freq_condition1}. In the case of motion along $\hat{\bf{y}}$ a larger $\bar{y}$ can be obtained by increasing the trapping power and lowering the NA. 
Assuming $\bar{y}\gtrsim\mathcal{O}(1~\text{mm})$ results in bounds on $T_y$ (from Eq.~\ref{eq:freq_condition1}) which exceed the residual gas temperature, i.e.~such that feedback damping would have to be larger than the minimum $\gamma_\text{min,y}$ given in Eq.~\ref{eq:NL_damping_condition} to avoid heating the gas.

Although Eq.~\ref{eq:freq_condition1} leads to relatively loose bounds on motion in the $\hat{\bf{z}}$ direction, stronger constraints on  these degrees of freedom can be derived from the fact that the sensor must not move outside the optical beams and that the scattered light needs to be collected by the lenses in order to be read out by the photodetectors. Assuming lenses of $\mathcal{O}(1~\text{cm})$ width, this is achieved provided the standard deviation of the translational motion of the sensor in the $\hat{\bf{y}}$-direction obeys $\sigma_y\ll\mathcal{O}(1~\text{mm})$. 
Using the definition of the standard deviation $\sigma_y$ as given in Eq.~\ref{eq:ang_mech_temp_variance}, this constraint can be  written as
\begin{align}\label{eq:freq_condition2}
    \sqrt{\frac{2k_BT_y}{\Omega_y}}y_0\ll 1\,\text{mm} ~,
\end{align}
where $y_0^2=1/(2m\Omega_y)$. 
Equation~\ref{eq:freq_condition2} can be further recast in the form of Eq.~\ref{eq:freq_condition1}, i.e.~as a frequency-dependent bound on temperature, from which lower bounds on damping rate $\gamma_y$ can be obtained and implemented on our setup. Although all the degrees of freedom of the sensor must respect both Eqs.~\ref{eq:freq_condition1} and \ref{eq:freq_condition2}, given that $\bar{y}$ is $\mathcal{O}(1~\text{mm})$ in the mechanical frequency range of interest, Eq.~\ref{eq:freq_condition2} is more stringent for this degree of freedom. In contrast, for motion in the angular direction $\hat{\bf\psi}$ and along $\hat{\bf{x}}$, given their more restricted values of $\bar{q}$, Eq.~\ref{eq:freq_condition1} is more constraining and thus what is used evaluate the maximum mechanical temperature and hence damping requirements for our system. We also note that both Eqs.~\ref{eq:freq_condition1} and \ref{eq:freq_condition2} are less stringent than the previous constraints discussed above for motion along the $\hat{\bf z}$-axis and in the angular directions $(\hat{\theta},\hat{\phi})$. Although a careful determination of the required values of $\gamma_q$ for each degree of freedom is left for future work, our estimates unambiguously show that the necessary damping rates to achieve stability of the trap, maintain a photon readout efficiency close to $\eta_D\approx 1$ and prevent additional backaction noise, are readily achievable with current damping techniques. 

Although not exploited in this work, translational motion along the $\hat{\bf{y}}$-axis and rotation around the $\hat{\bf{z}}$-axis lead to observable variation in the scattered optical field, and could therefore also be used for force or torque sensing. Leveraging these degrees of freedom could lead to further improvement of the detection capabilities  of our setup to dark matter-induced forces, as well as dark matter-induced torques, although this would require a different analysis to that provided in this section. We leave this to future work, and simply highlight the versatile use of our sensor to detect a wider variety of bosonic DM couplings.

Finally, despite controlling the mechanical temperature of all degrees of freedom, we still expect small losses of laser power due to random fluctuations (above $1 \sigma_q$) in the motion of the sensor or variations in the surface roughness. The latter are expected to dominate optical losses, and some of the scattered field is expected to escape into the vacuum chamber and heat the vacuum chamber walls. These will thus require cooling to maintain the temperature of the residual gas, sensor and associated apparatus. We estimate this additional cooling requirement  (beyond heat leaking in from the $300$~K exterior environment) to be $<\mathcal{O}(10~\text{mW})$ for our  0.2~mg sensor which is well within the $<\mathcal{O}(500~\text{mW})$ capacity of a pulse-tube liquid helium cryostat such as those employed in other low vibration setups \cite{Deshpande:2024bul}. Further, the losses for a 0.8~g sensor would increase to up to $\mathcal{O}(1~\text{W})$ due to the increased power requirements for the levitating beam, and for the trapping beams above 1~kHz. To avoid overwhelming the cooling power capabilities of our cryogenic system in this case, further attention will need to be paid to directing the horizontally scattered light out of the system, potentially via further shielding and/or photodetectors near the sensor.
We leave further discussion of these points (for both our 0.2~mg and 0.8~g sensors) to follow-up work in which more concrete details of the experimental setup will be provided (see also Ref.~\cite{Hamaide_2025}).

\section{Results\label{sec:results}}

We proceed to present projected exclusion bounds on the DM models introduced in Sec.~\ref{sec:dm_couplings}. To assess the sensitivity of our system to a DM signal of frequency $\omega_{\rm DM}$ we adopt the  signal-to-noise ratio (SNR), defined as 
\begin{align}\label{eq:SNR_1}
   \text{SNR}=\sqrt{\frac{\int_{\delta\omega} S_\text{FF}^{\rm DM}d\omega}{\int_{\delta\omega}S_\text{FF}^\text{noise}d\omega}}~,
\end{align} as our figure of merit, leaving a more complete likelihood analysis to future work. Here, $S_\text{FF}^{\rm DM}$ and $S_\text{FF}^\text{noise}$ denote the power spectra of the DM-induced force and background noise respectively (as a function of angular frequency) and $\delta \omega$ defines an angular frequency interval over which the power spectra are integrated.  We discuss the optimal location and width of this interval below. 

Whilst we have approximated the power spectra in Eq.~\ref{eq:SNR_1} as continuous functions of angular frequency, in practice their values are given in statistically independent bins of width $\Delta \omega = 2 \pi / T_{\rm int}$  (the frequency resolution of the experiment), where the integration time $T_{\rm int}$ is the total continuous time over which measurements of the sensor position are made for fixed experimental parameters $(\Omega,\bar{x})$. We emphasise that $T_\text{int}$ should not be confused with the total time required to scan over a range of frequencies while staying at the SQL at each frequency. The integrals of the signal and noise power spectra thus correspond to the $T_\text{\rm int}\rightarrow\infty$ limit of a sum over bin values times bin width. It follows that  $\delta \omega$ must be at least $\Delta \omega$. We now discuss the signal and noise power spectra which enter Eq.~\ref{eq:SNR_1} in turn.

The signal power spectrum $S_\text{FF}^{\rm DM}$ is formally defined to be the Fourier transform of the autocorrelation function of the inferred DM force on the sensor $\widetilde{F}$  (derived from measurements of its position relative to the support plate as discussed in Sec.~\ref{sec:setup}),  normalized such that\footnote{Our choice of normalization here differs from that of e.g.~Refs.~\cite{PhysRevX.4.021030,Arvanitaki:2014faa,Chaudhuri_2015,Manley:2020mjq} which use a $1/\sqrt{T_{\rm int}}$ normalization.  Both approaches leads to an identical expression for the SNR as the normalisation factors cancel.}
\begin{align}
    S_\text{FF}^{\rm DM}(\omega)=\lim_{T_{\rm int}\rightarrow\infty}\frac{1}{T_{\rm int}}\int_{0}^{T_{\rm int}} \langle \widetilde{F}(t)\widetilde{F}(0)\rangle e^{-i\omega t} dt ~.
\end{align}
For the quasi-monochromatic ULDM forces being considered, $S_\text{FF}^{\rm DM}$ consists of a prominent spike centered at $\omega_{\rm DM}$ of width $\Delta\omega_{\rm DM}\approx 10^{-6}\omega_{\rm DM} =  2\pi/T_{\text{coh}}$, deriving  from the DM's Maxwellian velocity distribution in the galactic halo. The majority of the DM signal is contained within this interval and we thus make the simplifying assumption that $S_\text{FF}^{\rm DM}\approx 0$ everywhere else. To ensure that the SNR captures the entirety of the signal it is necessary to set $\delta \omega \geq \Delta \omega_{\rm DM}$. 

\begin{figure*}[t!]
    \centering
    \includegraphics[width=1.8\columnwidth]{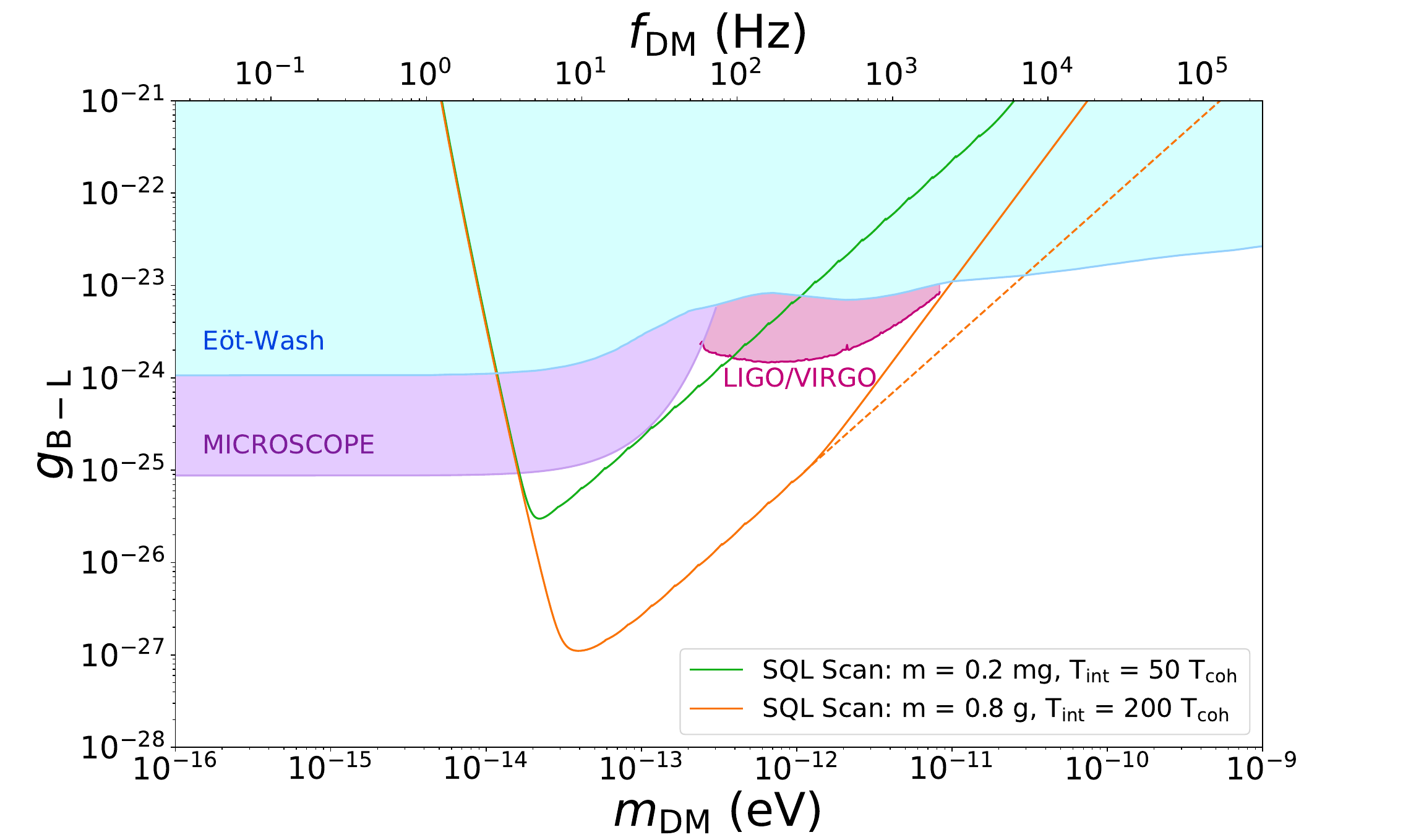}
    \caption{Projected sensitivies to $g_{\rm B-L}$ for our setup for sensors with mass 
    $m=0.2$~mg (green line), and $m=0.8$~g (orange), both with a four stage seismic isolation system. The solid lines indicate the reach assuming the maximum laser power of each of the $\hat{\bf x}$ directed lasers is $P_L = 70$~kW such that the in the case of the 0.8~g sensor the system is only tuned to reach SQL up to $f_{\rm DM} = 270$~Hz. The dashed line indicates the reach if sufficient power were available to reach SQL at every frequency. For the 0.2~mg sensor the SQL scan is performed in n = 20 bins, starting at frequency $f_1 = 5$ Hz. For the 0.8~g sensor we instead assume n = 10 bins and $f_1 = 10$ Hz.  For comparison to current experiments we show LIGO bounds, as well as bounds from fifth force searches with MICROSCOPE (purple) and E\"{o}t-Wash (blue). For further discussion of these bounds see Sec.~\ref{sec:results}.}
    \label{fig:bounds_B-L}
\end{figure*}

In our SNR we include both quantum and classical contributions to $S^{\rm noise}_{\rm FF}$, summing the various $S_{\rm FF}$'s discussed in Sec.~\ref{sec:noises}.  To minimise the quantum noises $S^{\rm q}_{\rm FF}$,  $\Omega$ and $\bar{x}$ should be tuned to satisfy the SQL condition in Eq.~\ref{eq:noisemin} for $\omega_{\rm DM}$. Formally, $S^{\rm q}_{\rm FF}$ is the sum of the total quantum noise from the horizontal beam ($S^{\rm tot}_{\rm FF}$ in Sec.~\ref{ssec:quantum_noises}) and the additional backaction (in the $\hat{\bf{x}}$ direction) from the vertical beam, which is independent of ($\Omega,\bar{x}$).    It is important to remember that achieving the SQL is a frequency-dependent statement. Having set ($\Omega,\bar{x}$) to optimize the quantum noises at the frequency $\omega_{\rm DM}$, the quantum noise at other frequencies $\omega \neq \omega_{\rm DM}$ is then above  SQL. In practice, the quantum noise spectrum that enters the SNR is \textit{flat} for $\omega\leq\omega_{\rm DM}$, scaling as $\propto\omega^4$ for $\omega\geq\omega_{\rm DM}$. For sufficiently small $\delta \omega$  however, it is reasonable to approximate the quantum noise in this interval as being flat with a value $S^{\rm q,min}_{\rm FF}(\omega_{\rm DM})$ which is the sum of the minimum horizontal quantum noises as defined in Eq.~\ref{eq:min_noise_SFF} with the constant contribution from the vertical beam.
A further implication of the frequency dependence of SQL is that for each different $\omega_{\rm DM}$ of interest, the experiment should ideally be reperformed with $(\Omega, \bar{x})$ tuned to satisfy Eq.~\ref{eq:noisemin} in each case. For now, we will focus on evaluating the sensitivity to a signal of frequency $\omega_{\rm DM}$ (assuming the system has been tuned to reach SQL at this frequency), before returning to discuss a realistic strategy with which to optimally probe a range of $\omega_{\rm DM}$ in due course.  

In order to maximise the SNR for a given $\omega_{\rm DM}$ we must judiciously select the frequency interval $\delta \omega$ over which to compare the signal and noise, in addition to $T_{\rm int}$ which governs the frequency resolution of the experiment.  If  $\Delta \omega \geq \Delta \omega_{\rm DM}$, the signal is entirely contained within one experimental frequency bin. This corresponds to the regime $T_{\rm int} \leq T_{\rm coh}$ in the time domain.  In this case, the selection $\delta \omega = \Delta \omega$ proves optimal: taking $\delta \omega $ any larger would contribute further to the denominator of Eq.~\ref{eq:SNR_1} without increasing the numerator. In this regime the SNR scales with  $\sqrt{T_{\rm int}}$. Now consider the case where $T_{\rm int} > T_{\rm coh}$  or equivalently $\Delta\omega <\Delta\omega_{\rm DM}$ such that the signal is spread across $N = T_{\rm int}/ T_{\rm coh}$ bins. The SNR computed in each individual bin is identical but we now have $N$ independent measurements of it.  The squares of the SNR in each bin can be added in quadrature to attain a total \textit{effective} SNR which benefits from a factor $N^{1/4}$ improvement relative to the result from each individual bin. The effective SNR in this regime scales as ($T_{\rm coh}T_{\rm int})^{1/4}$ (see e.g.~Ref.~\cite{PhysRevX.4.021030}). The relative sensitivity gain that can be achieved from further increasing the integration time when $T_{\rm int} > T_{\rm coh}$ is therefore more marginal than for when $T_{\rm int} \leq T_{\rm coh}$.

As we have seen, the quantum noises can only be optimised for a single signal frequency at a time. To search for DM over a wide range of values for $\omega_{\rm DM}$, ($\Omega, \bar{x}$) should ideally be tuned to reach SQL at each frequency, re-running the experiment for a period $T_{\rm int}$ each time. Given that $\omega_{\rm DM}$ is a continuous quantity, this would be unphysical. In practice, the DM frequency range of interest is divided into $n$ intervals of width $\epsilon$ whose lower endpoints occur at frequencies $\omega_i$ for $i = 1,...,n$. The experiment is then carried out $n$ separate times, each with a different ($\Omega, \bar{x}$) as necessary to satisfy the SQL condition at the frequency $\omega_i$.\footnote{Note that the choice to optimise at the lower endpoint of each frequency bin serves to minimize the deviation of the binned sensitivity from the sensitivity that would be obtained if the SQL could be reached continuously. This is due to the smaller relative scaling difference between noise at the SQL taken continuously over all $\omega$, and noise at frequencies above the SQL taken at a fixed (lower) frequency, compared to using noises at frequencies below the SQL taken at a fixed (higher) frequency. Nevertheless, the quantitative difference between these choices is small.}  In order to keep $n$ manageable, whilst ensuring that the quantum noise at each $\omega_{\rm DM}$ in the interval $\epsilon$ does not deviate from its SQL value (Eq.~\ref{eq:min_noise_SFF}) by more than an $\mathcal{O}(1)$ factor, we make the choice that  $\omega_{i+1}/\omega_{i}=\sqrt{2}$ such that $S^{\rm q, min}_\text{FF}(\omega_{i+1})/S^{\rm q, min}_\text{FF}(\omega_{i})=2$. This keeps the total number of coherent integration periods required to cover the frequency range $f_{\rm DM}\in[5~\text{Hz},5~\text{kHz}]$ to 20. Below this frequency range there is no benefit to tuning the system to SQL as seismic noise (which is independent of both $\Omega$ and  $\bar{x}$) becomes dominant. This choice of $\epsilon$ is sufficiently small such that the quantum noises in each $\epsilon$ interval  are approximately flat with value $S^{\rm q, min}_{\rm FF}(\omega_i)$. In essence this means that for each target frequency the SNR can be evaluated as above but by making the replacement
\begin{equation}
S^{\rm q, min}_{\rm FF} (\omega_{\rm DM}) = \begin{cases} 
 S^{\rm q, min}_{\rm FF} (\omega_{1})  & \textnormal{if} \hspace{1em}\omega_{\rm DM} < \omega_1 \\
  S^{\rm q, min}_{\rm FF} (\omega_{i}) & \textnormal{if } \hspace{1em} \omega_i \leq \omega_{\rm DM} < \omega_{i+1} \\
  S^{\rm q, min}_{\rm FF} (\omega_{n})  & \textnormal{if } \hspace{1em} \omega_{\rm DM} \geq \omega_n 
\end{cases}
\end{equation}. 

To maximize the parameter space that could be covered in a fixed total scan time (i.e.~including all $n$ iterations), we set the integration time for each run to be proportional to the DM coherence time at $\omega_i$ e.g.~$T_{\rm int}(\omega_i) = N T_{\rm coh}(\omega_i)$ with $N$  a frequency-independent constant. With this choice, 
the SNR for a DM signal of frequency $\omega_{\rm DM}\approx\omega_i$ can now be approximated as
\begin{align}
\label{eq:SNR}
    \text{SNR}\approx 
    \frac{0.1 m \tilde{g}} {m_n}\sqrt{\frac{2\rho_{\rm DM}\sqrt{N}}{S_\text{FF}^\text{noise}(\omega_i)\Delta\omega_{\rm DM}}} ~,
\end{align}
where as before $\tilde{g}$ is either $g_\text{B-L}$ or $y_nv$ and the factor of 0.1 corresponds to the difference in proton-to-nucleon ratio of the silica sensor and tungston-186 support plate. 
It is assumed that all of the noise contributions in $S_{\rm FF}^{\rm noise}$ are approximately flat over the region of integration $\Delta \omega_{\rm DM}$ given $\Delta \omega_{\rm DM}<\epsilon$, with the quantum noises taking the value $ \approx S^{\rm q,min}_{\rm FF}( \omega_i)$ (irrespective of the $\omega_{\rm DM}$ of interest) and the classical noises being evaluated at $\omega_{\rm DM}$. 

In our sensitivity forecasts we take  $N = 50$ for the 0.2~mg sensor. This corresponds to $T_\text{int}\approx 1\times 10^7$~s ($\sim 4$ months) for $f_{\rm DM}\approx 5$~Hz. With our choice of $\epsilon$ the resources required to scan the entire frequency range of interest are then  modest, at $\approx 11$ months. For the 0.8~g sensor we assume more significant resources at our disposal and instead show results for $N = 200$. Note that since the results for $\omega_{\rm DM}$ in different $\epsilon$ intervals come from independent experimental runs, our results for each of these intervals hold regardless of whether the other runs are carried out. Given practical time limitations, it may prove preferable to perform just some fraction of these runs (i.e.~to cover some fraction of the total scan interval) giving  priority to those which would provide access to the greatest amount of new parameter space.  The results presented here therefore serve as a guide to which intervals may prove most fruitful to cover at this integration time.

To allow for straightforward comparison with existing constraints and competing proposals working with this convention \cite{PhysRevX.4.021030,Arvanitaki:2014faa,Chaudhuri_2015,Manley:2020mjq}, we define our projected reach to be the parameter space in which the SNR~$\geq 1$. The minimum value of $g_\text{B-L}$ that could be detected at an SNR = 1 with our set up is 
\begin{align}\label{eq:bounds}
    g_\text{B-L}^{\rm min}= 6.9\times 10^{-25}\left(\frac{m_{\rm DM}}{10^{-13}~\text{eV}}\right)^{3/2}\left(\frac{0.2~\text{mg}}{m}\right)^{1/2}\left(\frac{N}{1}\right)^{1/4}~.
\end{align}
This scaling with DM mass holds for $m_{\rm DM}\gtrsim 2\times 10^{-14}$~eV for our 0.2~mg mass sensor but only for $m_{\rm DM}\gtrsim 4\times 10^{-14}$~eV for the 0.8~g mass sensor due to seismic noise dominating below this point.

In Fig.~\ref{fig:bounds_B-L} we show the projected sensitivity to $g_\text{B-L}$ at  SNR~$=1$ for $m=0.2$~mg (green) and for $m=0.8$~g (orange line), assuming the integration times stated above.  

For the 0.2~mg mass we assume that the SQL scan is performed in 20 bins, whose widths are set according the criterion above, with the lower endpoint of the first bin, $f_1$, at  5 Hz. For the 0.8~g mass we instead take $f_1 = 10$ Hz owing to the fact that seismic noise dominates below this frequency, rendering optimisation of the quantum noises largely redundant.   As expected, the sensitivity becomes limited by seismic noise at low frequencies, resulting in a rapid increase in the projected exclusion bounds. In the future some of this inaccessible parameter space could be recovered by placing the experiment underground where seismic noises are typically lower, or making use of differential measurements between two sensors of different materials, although as discussed in Sec.~\ref{sec:noises}  improvements to sensor fabrication and a lower residual gas pressure would also be required to reach the quantum floor. Without these improvements in pressure and fabrication, the total noise would be constant in this region and bounds would behave as $g_\text{B-L}\propto m_\text{DM}^{1/2}$ below $m_{\rm DM}\approx 2\times 10^{-14}$~eV (for a 0.2~mg sensor), due to the fact that since $N$ is fixed, the integration time is greater for lower DM masses.

For comparision, we show existing bounds on this coupling from E\"{o}t-Wash torsion balance experiments~\cite{Wagner_2012} and the MICROSCOPE satellite~\cite{Berg_2018}, both of which search for fifth forces which violate the Weak Equivalence Principle. For the latter we state the limit at the 90\% confidence level (CL), following the analysis presented in Ref.~\cite{amaral2024vectorwavedarkmatter} which takes into account the final experimental results reported in Ref.~\cite{MICROSCOPE:2022doy}.\footnote{Note that Ref.~\cite{Fayet:2025brx} also derives bounds on $g_{\rm B-L}$ from the final MICRSCOPE result using an alternative analysis. The resulting bounds are in agreement with the (corrected) result of Ref.~\cite{amaral2024vectorwavedarkmatter} (as shown in this work) in the low mass limit, but exhibit a different behaviour in the  high mass tail.} To correct for an erroneous conversion factor deployed in the limit derivation, we multiply the result presented in Ref.~\cite{amaral2024vectorwavedarkmatter} by a factor of 5/4. We also show bounds from LIGO  \cite{Pierce_2018,Abbott_2022,Guo_2019}. Our 0.2~mg sensor design outperforms existing experiments in the range $1.5 \times 10^{-14}~\text{eV}\lesssim m_{\rm DM}\lesssim 2\times 10^{-13}~\text{eV}$ by a factor of up to half an order of magnitude at $m_{\rm DM}\approx 2\times 10^{-14}$~eV. We note that experiments such as MICROSCOPE and E\"{o}t-Wash reach their peak sensitivity for DM-mediated forces with Compton wavelengths approximately larger than the Earth's radius. Below this point, their bounds are flat, while our projected sensitivity improves when going to lower DM mass as $g_\text{B-L}\propto m_{\rm DM}^{3/2}$, until becoming seismic noise-limited below $m_{\rm DM}\approx 2\times 10^{-14}$~eV. This scaling of $g_\text{B-L}$ is due to the fact that the optimal quantum noise which enters Eq.~\ref{eq:SNR} scales as $S_\text{FF}^\text {q,min}\propto m_{\rm DM}^2$ whereas  $\Delta \omega_{\rm DM}\propto m_{\rm DM}$. 

\begin{figure*}[t!]
    \centering
    \includegraphics[width=1.8\columnwidth]{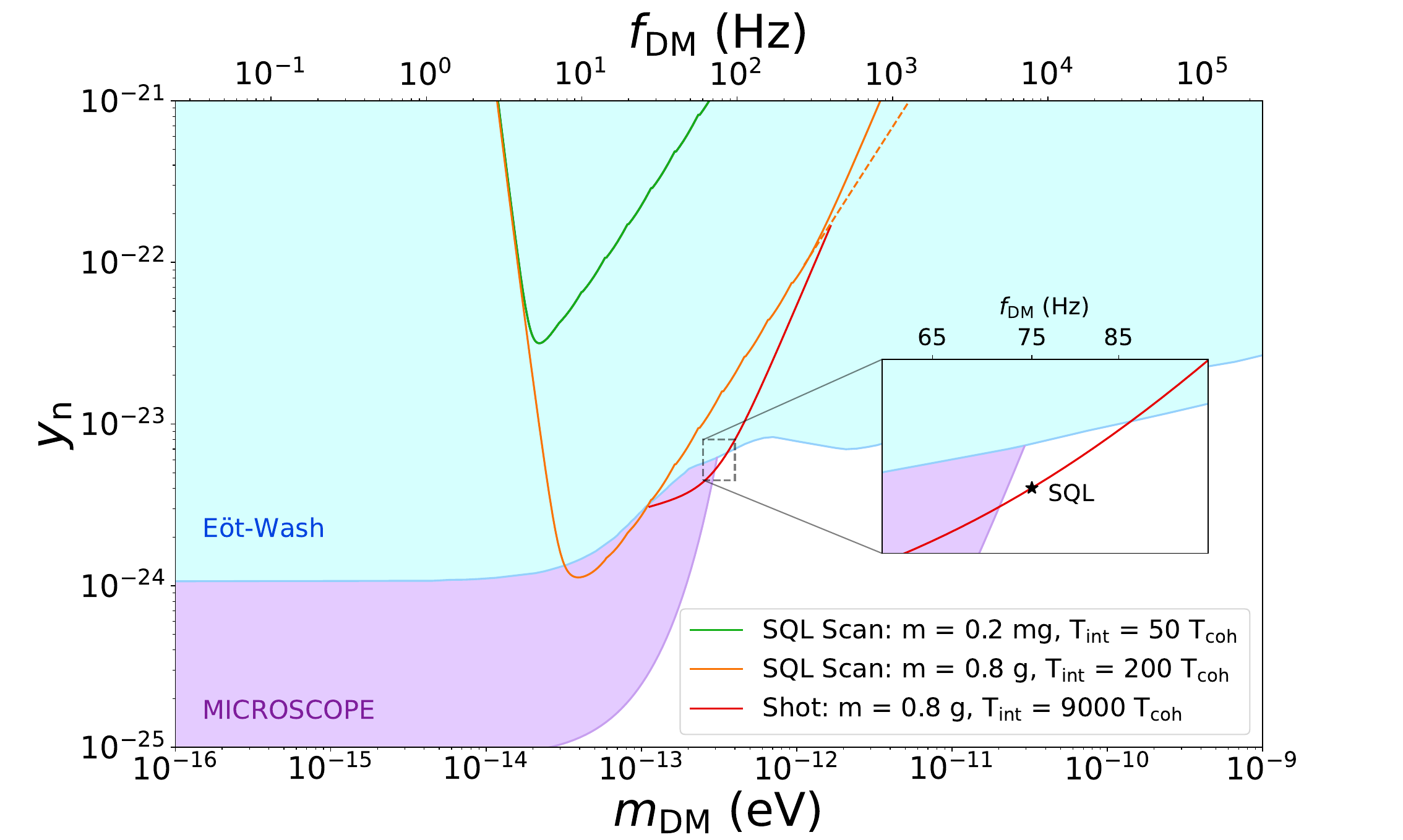}
    \caption{Projected sensitivies to $y_{n}$ for our setup for 
    $m=0.2$~mg (green line), and $m=0.8$~g (orange), both with a four stage seismic isolation system. The solid lines indicate the reach assuming the maximum laser power of each of the $\hat{\bf x}$ directed lasers is $P_L = 70$~kW such that for the 0.8~g sensor the system is only tuned to reach SQL up to $f_{\rm DM} = 270$~Hz. The dashed line  indicates the reach if sufficient power were available to reach SQL at every frequency.  For the 0.2~mg sensor the SQL scan is performed in n = 20 bins, starting at frequency $f_1 = 5$ Hz. For the 0.8~g sensor we instead assume n = 10 bins and $f_1 = 10$ Hz. We also show the projected sensitivity of a   `single shot' search in which the total measurement time available is devoted to a single run in which the system is tuned to SQL at 75 Hz. This allows new parameter space to be reached as emphasised in the inset. For comparison to current experiments we show bounds from fifth force searches with MICROSCOPE (purple) and E\"{o}t-Wash (blue). For further discussion of these bounds see Sec.~\ref{sec:results}.}
    \label{fig:bounds_scalar}
\end{figure*}

With the $m =0.8$~g  sensor we see an improvement in the sensitivity of $2 \times 10^{3/2}$ as a direct result of the increase in sensor mass (since $S_\text{FF}^{\rm DM}\propto m^2$ while $S_\text{FF}^\text{q, min}\propto m$ such that $g_\text{B-L}\propto 1/\sqrt{m}$).  The increased integration time at each frequency (by a factor of 4) provides a further improvement of $4^{1/4}$. This design would allow us to outperform current bounds over the entire region $1.5 \times 10^{-14}~\text{eV}\lesssim m_{\rm DM}\lesssim  1 \times 10^{-11}~\text{eV}$ by up to two orders of magnitude. Although in principle the sensitivity of the 0.8~g sensor would continue to scale as $g_\text{B-L}\propto m_\text{DM}^{3/2}$ indefinitely, above $m_\text{DM}= 10^{-12}$~eV the power requirements (on the $\hat{\bf x}$ directed beams) reach the limit of currently available (continuous wave) 1064~nm lasers. This contrasts to the 0.2~mg sensor where the total power requirement on the $\hat{\bf x}$-directed lasers is $2P_L\lesssim 5$~W 
across the frequency range where it outperforms existing experiments. To benchmark our 0.8~g sensor setup we assume that we have two $P_L\approx 70$~kW lasers, which are commonly used e.g.~for industrial applications, at our disposal. This would allow us to tune the system to reach the SQL up to $f\approx 270$~Hz. At frequencies above this (i.e.~for $m_{\rm DM} \gtrsim 1\times 10^{-12}$~eV), there is insufficient power to satisfy the SQL condition Eq.~\ref{eq:noise_condition_power_laser}. To search for DM signals in the frequency range $f\gtrsim 270$~Hz, we must therefore use the experimental run for which the quantum noises are optimized at $f\approx 270$~Hz, recalling that for target frequencies above this the quantum noises scale as $S_\text{FF}^\text{q}\propto\omega^4$, and exceed their SQL value. As a result of this cut-off, assuming that the bin widths for the SQL scan are set as described as above, the total number of bins required is 10. Further note that since we fix $T_{\rm int}$ for each experimental run to be a fixed multiple $N = 200$ of the DM coherence time at the frequency at which the SQL is attained, when searching for DM signals above this frequency, since $T_{\rm int}$ is fixed but $T_{\rm coh} \propto f_{\rm DM}^{-1}$, we have $N=200~(f/270~\text{Hz})$. As a result our forecasted exclusion bounds scale as  $g_\text{B-L}\propto m_\text{DM}^{9/4}$ above $f=270$~Hz.  However, assuming that larger laser powers can be obtained in the future, we also show with a dashed line the projected sensitivities which would be attained should it be possible to tune the system to reach the SQL at every frequency. Assuming a particular available laser power, the projected sensitivities scale as $g_\text{B-L}\propto m_\text{DM}^{3/2}$ up to the frequency at which the available power no longer allows us to tune our noise to be at the SQL (following Eq.~\ref{eq:noise_condition_power_laser}), above which point the sensitivity behaves as $g_\text{B-L}\propto m_\text{DM}^{9/4}$. As an illustration, should it be possible achieve  $P_L\approx 1~\text{MW}$, we may continue to operate our experiment at the SQL until $m_\text{DM}\approx 10^{-11}$~eV such that the projected bounds follow the dashed line up to this point. We also note that using higher optical frequencies may further improve our future reach, since the power requirement to reach the SQL in Eq.~\ref{eq:noise_condition_power_laser} goes as $P_L\propto 1/\omega_L$. However discussion of the implications of using a higher laser frequency on the availability of large laser powers, our choice of coating and substrate materials, as well as readout and cooling systems, are all left for future work.

Given our assumed laser power resources, Fig.~\ref{fig:bounds_B-L} shows that the projected sensitivity of our $m=0.8$~g sensor would improve on projected optical cavity sensitivities discussed in Refs.~\cite{Manley:2020mjq,Carney:2019cio,amaral2024vectorwavedarkmatter}, and is competitive with preliminary long-term projections of proposed magnetically levitated sensors such as POLONAISE \cite{Amaral:2024rbj}. We note however that long-term projections from POLONAISE only reach DM masses up to $m_\text{DM}=7\times 10^{-13}$~eV, whereas our 0.8~g sensor design could improve on current bounds up to $m_\text{DM}= 1 \times 10^{-11}~\text{eV}$. By searching at frequencies $f\gtrsim 5$~Hz our search for DM forces is also complementary to that of future long baseline atom interferometers, assuming they can deploy different isotopes \cite{Badurina:2019hst,Abe_2021}, and future torsion balance experiments. In particular, the sensitivity reach of our setup with a $m=0.8$~g mass is on par with preliminary forecasts for very long baseline atom interferometers  such as  MAGIS-100 (see Ref.~\cite{Abe_2021}) running with two different species, as well as future projections for torsion balance experiments \cite{Graham_2016}.

To highlight the novelty of our design we compare our sensitivity to that of  optically levitated nanoparticles which are more commonly studied than their more massive optically trapped counterparts. For the purpose of comparison we consider a (nearly) spherical silica nanoparticle of diamater 200~nm with mass $m\approx 10^{-17}$~kg as used e.g.~in Refs.~\cite{pontin2022directionalforcesensinglevitated,Jain:2016}, which is near the maximum size for a nanoparticle trapped at the optical focus using its dielectric properties (larger particles would no longer purely be in the Rayleigh scattering regime, with an expected behaviour closer to the geometric optics regime). We find the projected sensitivities of our 0.2~mg and 0.8~g sensors are $\mathcal{O}(10^5)$ and $\mathcal{O}(10^{6.5})$ better respectively than this reference nanoparticle. This behaviour can be explained as follows. The reference nanoparticle is expected to achieve $S_\text{FF}^\text{q, lev}\sim 10^{-43}$~N$^2/$Hz$~(f/1~\text{kHz})^2$, compared to $S_\text{FF}^\text{q, min}\sim 7.7\times 10^{-34}$~N$^2/$Hz$~(f/1~\text{kHz})^2$ for our 0.2~mg mass. As discussed in Sec.~\ref{ssec:quantum_noises}, once the sensor is large enough to begin scattering light in the geometric optics regime, the SNR is expected to scale as $\propto m^{1/2}$, in contrast to smaller sensors in the Rayleigh regime where the SNR remains invariant when scaling the mass. Our improvement in sensitivity is thus fully consistent with a scaling of  mass in the geometric optics regime by 10 orders of magnitude for the small mass, and 13 orders of magnitude for the large mass. Hence, although we have proposed a novel trapping method exploiting the comparatively less explored geometric optics regime and deploying masses up to 13 orders of magnitude higher than a typical levitated nanoparticle, we have shown our sensitivity to DM has nonetheless benefited from the scaling of the mass as $\propto m^{1/2}$ over 13 orders of magnitude, demonstrating a remarkable predictability in the behaviour of optomechanical sensors over such a large range of masses. 

Turning to the other end of the spectrum of optomechanical sensor masses, we now compare our sensitivities to those of a large setup like LIGO (from Refs.~\cite{Pierce_2018,Abbott_2022,Guo_2019}). Despite the fact that LIGO operates at SQL (off-resonance) at frequency $f\approx 100$~Hz with mirrors weighing $M=40$~kg each, which would naively result in a better sensitivity to DM by a factor of $\sqrt{2M/m}= 2\times10^4$, our sensitivity reach is comparable to LIGO's sensitivity at this frequency even with our smaller sensor mass of $0.2$~mg. Further, by tuning parameters of our setup to reach SQL at every frequency (in contrast to LIGO), we are able to improve on LIGO's bounds below 100~Hz such that near 5~Hz we can reach couplings up to an order of magnitude beyond  LIGO's lowest DM coupling sensitivity reach. This results from the fact that the observable at LIGO is a \textit{differential} measurement of the phase of the light in each arm at the end of the interferometry sequence. This depends on the relative position of its mirrors, as first shown in Ref.~\cite{Pierce_2018}. Such a measurement is only sensitive to the gradient of the DM induced force which is a factor of  $2L_\text{arm}/\lambda_\text{DM}\approx10^{-5}\,(m_\text{DM}/3\times 10^{-13}~\text{eV})$ smaller than the force itself, where $L_\text{arm}$ denotes the length of LIGO's arms. Oscillation of the DM force during a single roundtrip of the light in the arm can also lead to a relative phase, as shown in Ref.~\cite{Morisaki_2021}. In this case LIGO is sensitive to the amount of DM oscillation in the arm during the roundtrip time which is suppressed by $(m_\text{DM}L_\text{arm})^2\approx10^{-4}\,(m_\text{DM}/3\times 10^{-13}~\text{eV})^2$. This almost entirely negates LIGO's advantage to due to its larger mass, and emphasises the importance of targeted designs in optomechanical experiments such as MOLeQuTE, which can achieve greater sensitivities with less resources.

Note that to match experimental proposals targeting this model our sensitivity forecasts assume that the vector boson constitutes 100 \% of the local DM density. Should it only constitute a fraction, $\zeta$, the factor of $\rho_{\rm DM}$ in Eq.~\ref{eq:SNR} should be replaced with $\zeta \rho_{\rm DM}$. The impact of this is to increase the minimum detectable coupling $g_{\rm B-L}^{\rm min}$ by a factor of $1/\sqrt{\zeta}$.  In this way, our forecasts can simply be rescaled to assess the sensitivity to scenarios in which the B-L vector boson comprises only part of the dark sector.

Having discussed projected sensitivities of our setup to $g_\text{B-L}$, we now straightforwardly infer the sensitivity of our setup to a scalar DM coupling to neutrons as described in Sec.~\ref{sec:dm_couplings}. Recalling that the expressions for the B-L and scalar DM forces in Eqs.~\ref{eq:force_B-L} and \ref{eq:force_scalar} are the same up to the identification  $g_\text{B-L}\rightarrow v_{\rm DM} y_n$, the minimum observable value of $g_{\rm B-L}$ given in  Eq.~\ref{eq:bounds} may be simply rescaled by a factor $v_{\rm DM}^{-1}\approx 10^3$ to yield the sensitivity to $y_n$. This rescaling does not apply to the bounds from MICROSCOPE and E\"{o}t-Wash however, which yield the same value for $g_\text{B-L}$ and $y_\text{$n$}$ as they do not search for dark matter but for a fifth force dark mediator which does not experience a velocity suppression.\footnote{Note that the analysis of Ref.~\cite{Fayet:2025brx} explicitly accounts for the spin of the fifth force mediator in deriving bounds from the final MICROSCOPE results, finding a small difference between scalar and vector cases. } We compare these existing limits on $y_n$  to the projected sensitivities of our designs in Fig.~\ref{fig:bounds_scalar}. As shown, we find  that neither our 0.2~mg or 0.8~g sensor are sensitive to new regions of mass-coupling parameter space at the stated integration times.  

Noting the shape of our projections in relation to existing bounds, instead of spending time optimising the quantum noises at frequencies which are already tightly constrained by existing searches, it is advantageous to concentrate resources on those frequencies which are closest to reaching unchartered parameter space. The red line in Fig.~\ref{fig:bounds_scalar} shows the sensitivity that could be attained by performing a single run (a `single shot' search) with the system tuned to reach SQL at 75~Hz and an increased integration time of 3.5 years. For comparison with SQL scan type searches, we note that this corresponds to 9000 times the coherence time of a DM signal oscillating at the frequency to which the system is tuned to SQL i.e.~75~Hz. As a result, the sensitivity at 75~Hz is increased by a factor of 2.5 relative to the orange line, just entering new phenomenological territory as emphasised in the inset of Fig.~\ref{fig:bounds_scalar}. Note that since the system is only at SQL at 75~Hz, this improvement factor only applies at this frequency. Below 75~Hz the sensitivity curve becomes less steep than that achieved with an SQL scan, scaling with DM mass as $y_n \propto m_\text{DM}^{1/4}$ as the system is above SQL. Since, as discussed in Sec.~\ref{sec:noises} for frequencies below the SQL frequency, the total quantum noise of the system remains constant, this scaling results from the increase in the signal coherence time (which enters the SNR via  $(T_\text{coh}T_\text{int})^{1/4}$) at smaller masses.  Note that the projections presented in this work  could be substantially improved by using an array of $P$ sensors, which, assuming they have not been prepared in an entangled state, would improve the sensitivity by a factor $\sqrt{P}$, thus reducing the integration time requirements to reach new parameter space.


\section{Conclusion}
Several well-motivated models of vector and scalar ULDM are predicted to give rise to feeble oscillatory forces on SM particles. In this work, we propose a novel optomechanical setup consisting of a high mass optically trapped sensor, specifically designed to detect these forces. 

The foundations of our design are informed by a systematic analysis of the signal-to-noise ratio which we adopt as our figure of merit. We demonstrate that, for ULDM force-sensing, it is preferable to levitate larger objects whose laser interactions fall within the geometric optics regime, as opposed to Rayleigh scattering nanoparticles which are more routinely deployed in optically trapped systems.  Whereas the quantum noises relevant to trapped nanoparticles scale identically with sensor mass to the DM signal, this identity breaks upon entering the geometric optics regime such that increasing the sensor mass leads to an improved SNR, provided classical noises remain under control.

We further provide the first formal, consistent optimisation of the quantum noises intrinsic to optically trapped systems with respect to the experimental degrees of freedom, carefully taking into account the relationship between the mechanical frequency of the trap and the readout power. This detail had been glossed over in previous treatements and is important in both the Rayleigh and geometric optics regimes. Applying this analysis to our system, we find that the minimum quantum noise that can be achieved for each target frequency, within the range of experimentally feasible parameter values, occurs off resonance. Informed by these findings, we propose a feasible setup involving a large, optically trapped sensor of mass 0.2~mg whose weight is supported by a vertical beam. By considering the relevant classical noise sources we show that, relying only upon modest resources and off-the-shelf technologies, our experiment is capable of operating in the quantum-noise limited (SQL) regime down to frequencies of $f_{\rm DM}\approx 5$~Hz whereupon seismic noise becomes dominant, unless additional techniques (e.g.~a differential measurement involving two sensors of different compositions) are applied to further suppress this.

In the short term, the projected sensitivity of this experiment to vector B-L coupled ULDM surpasses fifth-force tests and LIGO searches for DM  masses in the range $m_{\rm DM}\in[1.5\times 10^{-14},2\times 10^{-13}]$~eV by up to half an order of magnitude. Further improvements to the reach could be achieved in the future both by upscaling the sensor mass to $\sim 0.8$~g (provided the necessary trapping power can be obtained) and extending our modest integration time. Not only does this open the window to smaller B-L couplings, but it increases the range of DM masses over which our experiment outperforms existing searches. Our 0.8~g mass design is also capable of reaching unexplored parameter space for scalar ULDM coupling to neutron (or proton) number, provided resources are concentrated on a single run tuned to reach SQL at 75 Hz. 

These promising forecasts surpass those presented in theoretical studies of suspended optical cavities \cite{Carney:2019cio,amaral2024vectorwavedarkmatter} which deploy similar sensor masses, and are often used as the benchmarks for optomechanical DM detection, as well as projected sensitivities based on predicted noise in optically levitated nanoparticles \cite{Gosling:2023lgh}. They are also on a par with, or exceed predictions for proposed magnetically levitated detectors \cite{Li:2023wcb,Kalia:2024eml}, including the long-term projected sensitivities of POLONAISE in Ref.~\cite{Amaral:2024rbj}. Given that such systems typically involve the levitation of higher mass sensors, this result is largely a consequence of the ability of our experiment to reach the quantum noise floor with existing technologies.

Several directions to improve the performance of our proposed setup exist. Lower noise could be achieved by harnessing optically squeezed light, such as that used at LIGO \cite{PhysRevX.13.041021}, or by use of an array of identically trapped sensors. The latter could be further improved by entangling their states such that the sensitivity would improve linearly with the number of sensors \cite{Brady_2023}. Further, the proposed experiment design possesses several features which have not been exploited in the current work, with a more  detailed treatment left to future studies. These include the possibility of leveraging the directionality of the setup to simultaneously monitor the position of the particle in both the $\hat{\bf{x}}$ and $\hat{\bf{y}}$ directions, thus opening the possibility of seeing the DM wind in the event of a positive detection. Simultaneously detecting forces in two directions at the same frequency would also increase the statistical likelihood of a DM signal, thus improving the projected sensitivities presented  in this work by a factor of $\sqrt{2}$. Alternatively, to double the scan rate across DM frequencies, we may tune each direction to detect forces at different frequencies. We also leave the torque sensing capabilities of this system (around the vertical $\hat{\bf{z}}$-axis), which could offer the means to probe an increased compendium of ULDM couplings such as pseudoscalar DM coupling to spin, to future work.

A top priority for ULDM force-sensing is a fully tunable system which is able to capitalise on the enhancement in sensitivity which arises when the SQL can be obtained on resonance. Whilst, due to restrictions coming from the Rayleigh range $x_r$, our system is ultimately prevented from realising this, it does achieve SQL closer to resonance than previously proposed designs. Not only does this study therefore mark a crucial step forwards in the optimisation of optomechanical systems for force sensing, but it is hoped that the novel analysis and resulting conclusions in this work may offer concrete insight into the most fruitful experimental directions to pursue in the future.

\begin{acknowledgments}
We are grateful for helpful discussions with Jonathan Gosling, Markus Rademacher, Felipe Almeida da Silva, Daniel Carney, David Moore, Thomas Penny, Marko Toros, Felix Yu, Gianpiero Mangano, Josh Ruderman and Ken Van Tilburg. 
We thank Giacomo Marocco, Sebastian Ellis and Pierre Fayet for helpful comments on an early draft of this paper.
This work was partially supported by the research grant number 2022E2J4RK "PANTHEON: Perspectives in Astroparticle and Neutrino THEory with Old and New messengers" under the program PRIN 2022 (Mission 4, Component 1,
CUP I53D23001110006) funded by the Italian Ministero dell’Universit\`a e della Ricerca (MUR) and by the European Union – Next Generation EU. L.H.~ackowledges support from STFC grant ST/W006170/1 and the Alexander von Humboldt Foundation. 
P.B.~acknowledges support from STFC grant ST/W006170/1 and EPSRC grants EP/W029626/1, EP/N031105/1. 
H.B.~acknowledges partial support from the STFC HEP Theory Consolidated grants ST/T000694/1 and ST/X000664/1.  A.G.~acknowledges support from NSF grants PHY-2409472 and PHY-2111544, DARPA, the John Templeton Foundation, the W.M.~Keck Foundation, the Gordon and Betty Moore Foundation Grant GBMF12328, DOI 10.37807/GBMF12328, the Alfred P.~Sloan Foundation under Grant No.~G-2023-21130, and the Simons Foundation.
\end{acknowledgments}

\appendix

\section{Trapping potential}\label{app:trapping_potential}
We work in SI units in all the appendices. In this appendix we derive the expression for the harmonic oscillator trap in the $\hat{\bf{x}}$-direction which is created with the radiation pressure from the $\pm\hat{\bf{x}}$-beams. We remind ourselves the average (equilibrium) distance between each  surface of the sensor and the nearest focus is $\bar{x}$, while $x_r=\lambda_L/(\pi\text{NA}^2)\gtrsim \lambda_L/2$ is the Rayleigh range of the beams. Supposing the sensor is slightly displaced from equilibrium by $x\,\hat{\bf x}$ (where $\langle x\rangle=0$ and $\abs{x}/\bar{x}\ll 1$), the two surfaces are now at different distances $\bar{x}\pm x$ from the nearest focus (see Fig.~\ref{fig:setup_diagram} and Sec.~\ref{sec:setup} for further details). Provided $\bar{x}\gg x_r$ and $\bar{x}>h/\text{NA}$, where $h$ is the height of the sensor, the intensity of the beam on either surface in equilibrium is $I(\bar{x}) = \frac{I_0 x_r}{\bar{x}}$ while at a position $(\bar{x}\pm x)$ this expression can be re-written to first order in $x$ as
\begin{align}
I_\pm(x) = \frac{I_0 x_r}{(\bar{x} \pm x)}
\approx \frac{I_0 x_r}{\bar{x}}\left(1 \mp \frac{x}{\bar{x}}\right)~.
\end{align}
 The total scattered power (i.e.~from both surfaces)  is $P_{\textnormal{scatt}}=2I_0x_rA/\bar{x}$, where $A$ is the cross-sectional (surface) area of the sensor normal to the $\hat{\bf x}$ direction. We deploy this in the quantum noise calculations detailed in Sec.~\ref{ssec:quantum_noises}.  The scattering force on each surface (denoted $+/-$) is 
\begin{equation}
F_{\textnormal{scatt,+/-}} = \pm\frac{2I_0 x_rA}{\bar{x} c}\left(1 \mp \frac{x}{\bar{x}}\right) \mathbf{\hat{x}} ~.
\end{equation}
The net force on the sensor from both beams is thus
\begin{align}
F_{\textnormal{scatt}}= - \frac{4I_0 x_rA}{\bar{x}^2 c}x  \,\mathbf{\hat{x}}\, ~,
\end{align}
which corresponds to the force on a sensor in a harmonic oscillator in the $\hat{\bf x}$-direction, as described in Sec.~\ref{sec:noises}.

\section{Quantum Optical Noises}\label{app:noises_intro}

Optically trapped setups rely on the coupling between an object (sensor) and an optical field (laser) such that the object is in a simple harmonic oscillator (trap). The Hamiltonian of the system takes the form \cite{Aspelmeyer:2013lha,Carney:2019cio}
\begin{align}\label{eq:OM_hamiltonian}
    \hat{H}=\hbar\omega_L\hat{a}^\dagger\hat{a}+\hbar\Omega\hat{b}^\dagger\hat{b}+\frac{G}{x_0}\,x\,X ~,
\end{align}
where $(\hat{a}^\dagger,\hat{a})$ and $(\hat{b}^\dagger,\hat{b})$ are the creation/annihilation operators for the optical field and mechanical trap modes respectively, and we focus here on the $\bf{\hat{x}}$-direction (i.e.~$\Omega=\Omega_x$). In the last term we define $x=(\hat{b}+\hat{b}^\dagger)/\sqrt{2}$ as the position of the sensor, $X=(\hat{a}+\hat{a}^\dagger)/\sqrt{2}$ as the amplitude quadrature of the optical field, and $G$ as the optomechanical coupling. The latter is a function of the laser power and depends on the details of the optomechanical setup in question.

Quantum measurement noises arise as an inherent byproduct of the coupling between the trapped sensor and the laser. Irreducible fluctuations of the amplitude of optical field (i.e.~photons) used to trap and read out the motion of the sensor, scatter off the sensor and excite mechanical motion. 
The mechanical excitation rate i.e the rate at which motion (phonons) is excited (see Refs.~\cite{Clerk_2010,Magrini:2020agy} for more details), is
\begin{align}
    \nonumber \Gamma_\text{recoil} &= \bar{\dot{N}}\left(\frac{p_L}{p_0}\right)^2 \\ &= \frac{2P_\text{scatt}}{mc^2}\frac{\omega_L}{\Omega} ~,
\end{align}
where $m$ is the sensor's mass, $P_\text{scatt}=\bar{\dot{N}}\hbar\omega_L\propto P_L\propto G^2$ is the scattered power (with $P_L$ is the laser power), and the final numerical factor may further include a small correction due to the Gouy phase shift (see Refs.~\cite{PhysRevA.102.033505,Tebbenjohanns_2019} for details). Here $\bar{\dot{N}}$ is the average photon rate, while $(p_L,\omega_L)$ are respectively the momentum and angular frequency of the photons, and $p_0^2=\hbar^2/(4x_0^2)$ is the zero-point momentum, where again $x_0^2=\hbar/(2m\Omega)$. The zero-point momentum in a harmonic potential is related to the momentum variance analogously to the case of the position variance, i.e.~such that $\langle p^2\rangle=(2n+1)\,p_0^2$. The rate of momentum deposition per unit frequency in the trap has the power spectrum
\begin{align}\label{eq:def_backaction_noise}
    S_\text{FF}^\text{ba}=\hbar^2\frac{\Gamma_\text{recoil}}{x_0^2} ~.
\end{align}
where `ba' stands for backaction noise.

Further, the position of the particle is estimated using the optical phase. Fluctuations of the optical phase therefore lead to uncertainty in the sensor's position. Although these fluctuations are  a white noise, the phase uncertainty $\Delta\Theta$ is improved by averaging over the number of measured photons $\bar{\dot{N}}_\text{meas}=\eta_D\bar{\dot{N}}$. Following Ref.~\cite{Clerk_2010} we identify $(\Delta\Theta)^2=1/(4\bar{\dot{N}}_\text{meas})$, from which the power spectrum of the estimated position - called imprecision noise - is given by
\begin{align}\label{eq:def_shot_noise}
    \nonumber S_{\rm xx}^\text{imp}&= \frac{1}{16\,\bar{\dot{N}}_\text{meas}\,k_L^2}  \\ &=\frac{x_0^2}{4\,\Gamma_\text{meas}} ~. 
\end{align}
As discussed in Sec.~\ref{sec:noises} we assume $\Gamma_\text{meas}\approx\Gamma_\text{recoil}$ (i.e.~$\eta_D\approx 1$).  Equations.~\ref{eq:def_backaction_noise} \& \ref{eq:def_shot_noise} can be combined to yield $S_\text{FF}^\text{ba}S_{\rm xx}^\text{imp}=\hbar^2/4$. This saturates the uncertainty principle as expected since the backaction represents the variance of momentum deposited on the sensor while the imprecision noise represents the variance in the (estimated) position.
The backaction power spectrum can be rewritten in terms of position for easier comparison between the two. Fourier transforming the equation of motion of the harmonic oscillator, including a damping rate (e.g.~feedback damping or thermal damping, see Sec.~\ref{ssec:th_noises}) and a driving force $F(t)$ (e.g.~from $GxX$, a coupling to DM, or a thermal bath) yields
\begin{align}
    x(\omega)=\chi(\omega)F(\omega) \,\,\,\rightarrow \,\,\, S_\text{\rm xx}(\omega)=\abs{\chi(\omega)}^2 S_\text{FF}(\omega) ~,
\end{align}
where $(x(\omega),F(\omega))$ are the Fourier amplitudes and $\chi(\omega)=m^{-1}(\Omega^2-\omega^2+i\gamma\omega)^{-1}$. 
When detecting an external force acting on the sensor, $S_{\rm xx}$ is ultimately what is directly tied to the observable $x$, whereas the forces $F$ and thus associated $S_{\rm FF}$ are inferred.  $S_{\rm xx}$ therefore allows accurate comparison of the motion of several modes of the sensor including vibrational modes, as discussed in Sec.~\ref{ssec:th_noises}.

$\\$
\section{Angular trapping and control}\label{app:angular_trap}

As discussed in Sec.~\ref{ssec:trap_stability}, our sensor can be kept from rotating beyond the point where it significantly impacts the sensor's reconstructed position or induces backaction noises by using radiation pressure of the various focused beams to make angular traps and controlling the angular harmonic oscillator energy level of the sensor in each direction. In our case, this can be understood as the restoring torque on the sensor which appears when rotating since one side of the sensor's surface which is facing the beam gets closer to the focus of a beam while the other gets further. The side which gets closer to the focus experiences higher radiation pressure while the other side experiences lower radiation pressure, leading to a restoring net torque which creates the desired angular trap. This trap has an analogous Hamiltonian to the translational motion trap (as shown explicitly for $x$ in App.~\ref{app:trapping_potential}). Taking for illustration the rotations around the $\hat{\bf{y}}$-axis (i.e.~in the $\hat{\bf{\phi}}$-direction) , we write
\begin{align}
    \hat{H}=\frac{\hat{L}_y^2}{2I}+\frac{1}{2}I\,\Omega_\phi^2\hat{\bf\phi}^2 ~,
\end{align}
where $I$ is the sensor's moment of inertia around the $\hat{\bf{y}}$-axis, $\hat{\phi}$ here is the $\phi$-direction coordinate operator, and $\hat{L}_y$ is the angular momentum around the $\hat{\bf{y}}$-axis. The associated mechanical frequency is obtained by calculating the torque on the sensor $\tau=I\ddot{\phi}$. Explicitly we have (dropping operator hats): 
\begin{align}\label{eq:angular_trap}
   \tau = I\ddot{\phi}=-\frac{4L^3WI_0}{3w_z}\phi\,\,\,\,\rightarrow\,\,\,\, \Omega_\phi = \sqrt{\frac{4L^3WI_0}{3Iw_z}} ~,
\end{align}
where $w_z$ is the vertical beam's waist and $(W,L)$ are the width and length of the sensor respectively. In our case $L=W$ for both the 0.2~mg and 0.8~g sensors. Note that Eq.~\ref{eq:angular_trap} can be straightforwardly adapted to all Euler rotational directions choosing the appropriate sensor dimensions among $(h,W,L)$ and beam waists $(w_x,w_y,w_z)$ which refer to the beams along the $(\hat{\bf{x}},\hat{\bf{y}},\hat{\bf{z}})$ axes respectively. The associated zero-point angle of the angular motion in the $\hat{\bf\phi}$ direction is given by
\begin{align}\label{eq:angle_zpf}
    \phi_0=\sqrt{\frac{\hbar}{I\Omega_\phi}} ~.
\end{align}
Given that the sensor is much larger than the wavelength of light, we can assume that the fluctuations of the optical field which scatter along the length and width of the sensor are independent and thus exert a independant torque noise proportional to the moment arm  (distance from the axis running through the middle of the face of the sensor). Given further that the backaction force noise on the sensor is homogeneous at every position along the sensor's face with average moment arm length $L/4$, the torque backaction noise can be expressed as
\begin{align}\label{eq:torque_noise}
    S_{\tau\tau}=S_\text{FF}\,\left(\frac{L}{4}\right)^2 = \frac{\Gamma_\phi}{\phi_0^2} ~.
\end{align}
Using the second equality along with \ref{eq:angle_zpf} we may obtain the torque excitation rate $\Gamma_\phi$, assuming $S_\text{FF}=(2/5)\,\omega_LP_\text{scatt}$ (as discussed in Sec.~\ref{sec:noises}) where $P_{\rm scatt}$ is the total scattered power, which for motion in the $\hat{\bf\phi}$ direction corresponds to the power of the vertical beam. To use Eq.~\ref{eq:torque_noise} to put bounds on $\gamma_\phi$, we begin by recalling that the variance of $\phi$ is given by $\langle\phi^2\rangle=(2n_\phi+1)\,\phi_0^2$. We then impose an upper bound on the standard deviation of $\phi$, $\sigma_{\phi} = \sqrt{\langle\phi^2\rangle}$ at the value beyond which this motion would introduce an excessive noise on the inferred position of the sensor (in this case, $\sigma_{\phi,\text{rough}}$). This condition thus translates to an upper bound on the phonon number $n_\phi$. Using the steady state relationship $n_\phi=\Gamma_\phi/\gamma_\phi$ as discussed in Sec.~\ref{ssec:trap_stability} (also see Refs.~\cite{Millen_2020,Jain:2016}), we can then derive a lower bound on the angular motion damping rate $\gamma_\phi$ required to keep $\sigma_{\phi}$ within the aforementioned limit.

%

\end{document}